%% file: Main.tex
\documentclass[letterpaper, 12pt, one column]{article}
\usepackage{multicol}
\usepackage{fullpage}
\usepackage{multirow}
\usepackage{threeparttable} % enabling footnotes in tables.
\makeatletter \g@addto@macro\TPT@defaults{\footnotesize} \makeatother
\usepackage[font=small,labelfont=bf]{caption}

\usepackage{titling}
\newcommand{\subtitle}[1]{%
  \posttitle{%
    \vskip0.5em
    \par\end{center}
    \begin{center}\large#1\end{center}
    \vskip0.5em}%
}

\newcommand{\authornote}[1]{%
  \preauthor{%
    \vskip2em
    \begin{center}
    \par\end{center}
    \begin{center}\large#1\end{center}
    \vskip0.5em}%
    \postauthor{}
}

\usepackage[english]{babel}
\usepackage[T1]{fontenc}
\usepackage{color}
\usepackage{soul}
\usepackage{graphicx,wrapfig} %allows you to use jpg or png images. PDF is still preferred %wrapfig is alsoa nice feature
\usepackage{subfig}
\usepackage{float} %allows to fix the position of the figure - sometimes needed
\usepackage[rightcaption]{sidecap}
\usepackage[section]{placeins}
\sidecaptionvpos{figure}{c}

\usepackage[colorlinks = true,
            linkcolor = blue,
            urlcolor  = blue,
            citecolor = blue,
            anchorcolor = green]{hyperref}

\usepackage{lineno}
\usepackage{comment}
\usepackage{cite}
\usepackage{authblk}%author package

\newcommand{\snn}{\ensuremath{\sqrt{s_\mathrm{_{NN}}}}}

\usepackage{xspace}	% Include xspace

\newcommand{\pp} {$pp$}

\def \be {\begin{equation}}
\def \ee {\end{equation}} 
\def \bea {\begin{eqnarray}}
\def \eea {\end{eqnarray}}

\title{\textbf{The RHIC Cold QCD Program}}

\subtitle{\textbf{White Paper} \\ \vspace{1cm}

\textit{contribution to} \\ \textbf{the NSAC Long-Range Planning process}}

\authornote{\textbf{\underline{Authors for the RHIC SPIN Collaboration}}\footnote{The \textbf{RHIC Spin Collaboration} consists of the spin working groups of the RHIC collaborations, many theorists and members of the BNL accelerator department.}}

\author[1]{Elke-Caroline Aschenauer}
\author[2]{Kenneth Barish}
\author[1]{Alexander Bazilevsky}
\author[1]{Xiaoxuan Chu}
\author[3]{James Drachenberg}
\author[1]{Oleg Eyser}
\author[4]{Renee Fatemi}
\author[5]{Carl Gagliardi}
\author[6]{Sanghwa Park}
\author[1]{Vincent Schoefer}
\author[7]{Ralf Seidl}
\author[8]{Scott Wissink}
\author[9]{Qinghua Xu}
\author[10]{Maria Zurek}

\affil[1]{\textit{Brookhaven National Laboratory, Upton, New York 11973}}
\affil[2]{\textit{University of California, Riverside, California 92521}}
\affil[3]{\textit{Abilene Christian University, Abilene, Texas 79699}}
\affil[4]{\textit{University of Kentucky, Lexington, Kentucky 40506-0055}}
\affil[5]{\textit{Texas A\&M University, College Station, Texas 77843}}
\affil[6]{\textit{Mississippi State University, Mississippi State, Mississippi 39762}}
\affil[7]{\textit{RIKEN Nishina Center for Accelerator-Based Science, Wako, Saitama 351-0198}}
\affil[8]{\textit{Indiana University, Bloomington, Indiana 47408}}
\affil[9]{\textit{Shandong University, Qingdao, Shandong 266237}}
\affil[10]{\textit{Argonne National Laboratory, Lemont, Illinois 60439}}

\begin{document}
\setcounter{tocdepth}{3}
\date{}
\maketitle
\thispagestyle{empty}
%\linenumbers

\pagenumbering{roman}

\newpage
\tableofcontents
\pagenumbering{arabic}

\newpage
\section{Executive Summary}
\input{ExecutiveSummary}

\clearpage
\newpage
\section{Polarized High-Energy Proton Beams}
\input{HadronBeams}

\newpage
\section{Collinear Proton Structure}
\input{Collinear}

\newpage
\section{Three-dimensional Structure}
\input{3DImage}

\newpage
\section{Appendix}
\input{Appendix}

\newpage
\bibliographystyle{h-elsevier}
\bibliography{ref.bib}
\end{document}

%% file: ExecutiveSummary.tex
RHIC has produced a remarkable breadth of physics results over the years, with compelling discoveries in both Hot and Cold QCD. It is critical to utilize the last three years of operations to complete the extraordinarily rich program that is uniquely possible with the $pp$, $pA$ and $AA$ collisions provided by RHIC. 

A significant piece of the RHIC legacy is the 25 years of innovation in accelerator science and experimental techniques necessary to collide highly polarized, high-energy proton beams. These achievements, discussed in detail in Section~\ref{HadronBeamsChapter}, include the design and construction of the the world's highest luminosity polarized proton source, the use of Siberian snakes to reduce the depolarizing effects of the resonance field harmonics, %the ability to set bunch-by-bunch polarization directions in order to minimize the systematic effects due to correlations between the spin direction and bunch intensity, 
the implementation of spin rotators to provide  proton beams polarized in the longitudinal, transverse or radial direction and the development of techniques to maintain orbit and emittance stability from injection to full energy in order to maximize polarization lifetimes. In parallel, new techniques and tools were developed to monitor and evaluate the quality of the beams. As a result it is now possible to make precision measurements of the beam spin tune, to extract the transverse/radial polarization component in a longitudinally polarized beam and precisely measure the spin-dependent relative luminosities. These advances, along with the design, construction, and operation of absolute and relative high precision hadron polarimetry have played an essential role in the success of the RHIC experimental Cold QCD program and have laid the foundation for the design of the future Electron-Ion Collider's (EIC) highly polarized high energy hadron beams.

\begin{figure*}[tbh]
\centering
\includegraphics[width=0.45\textwidth]{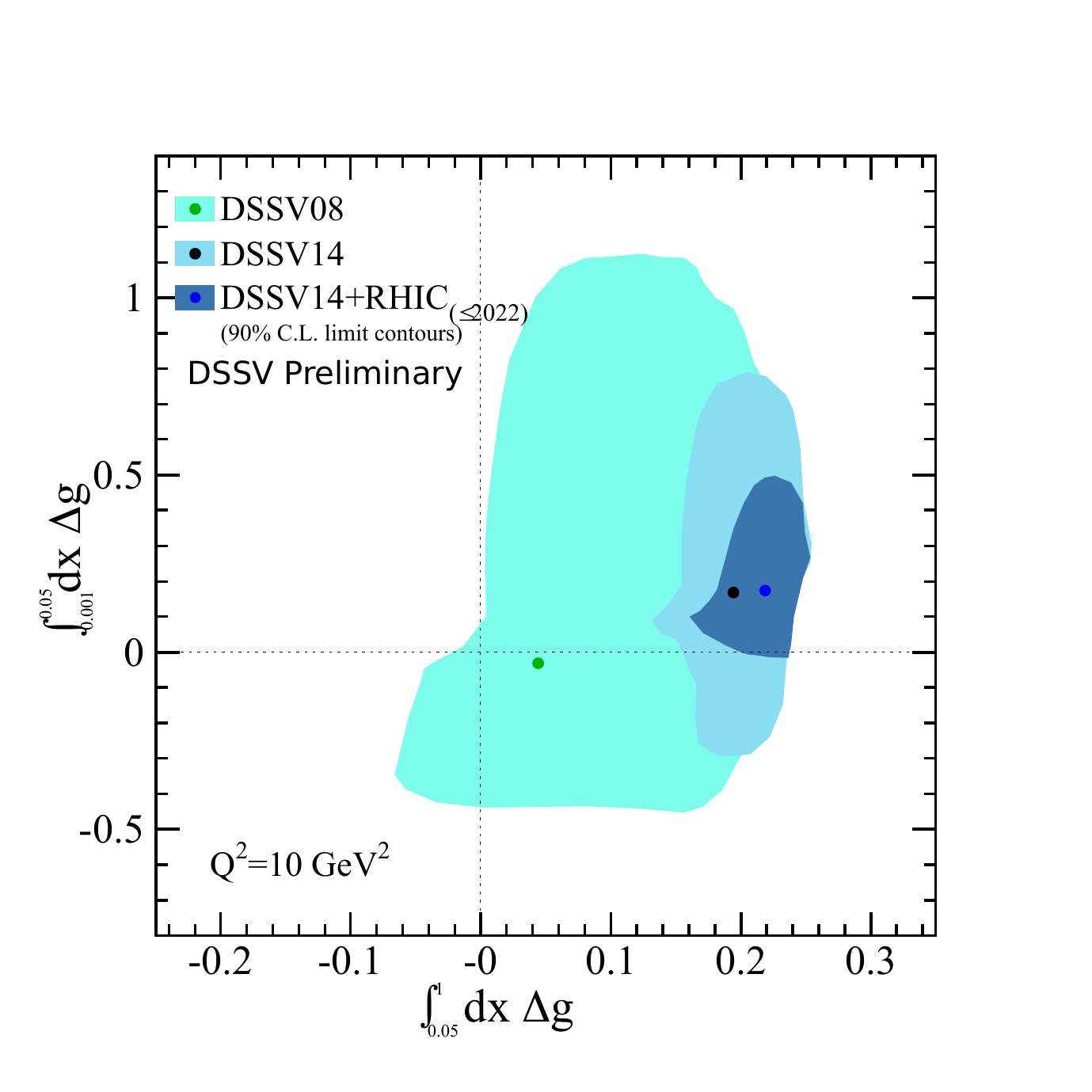}
\includegraphics[width=0.53\textwidth]{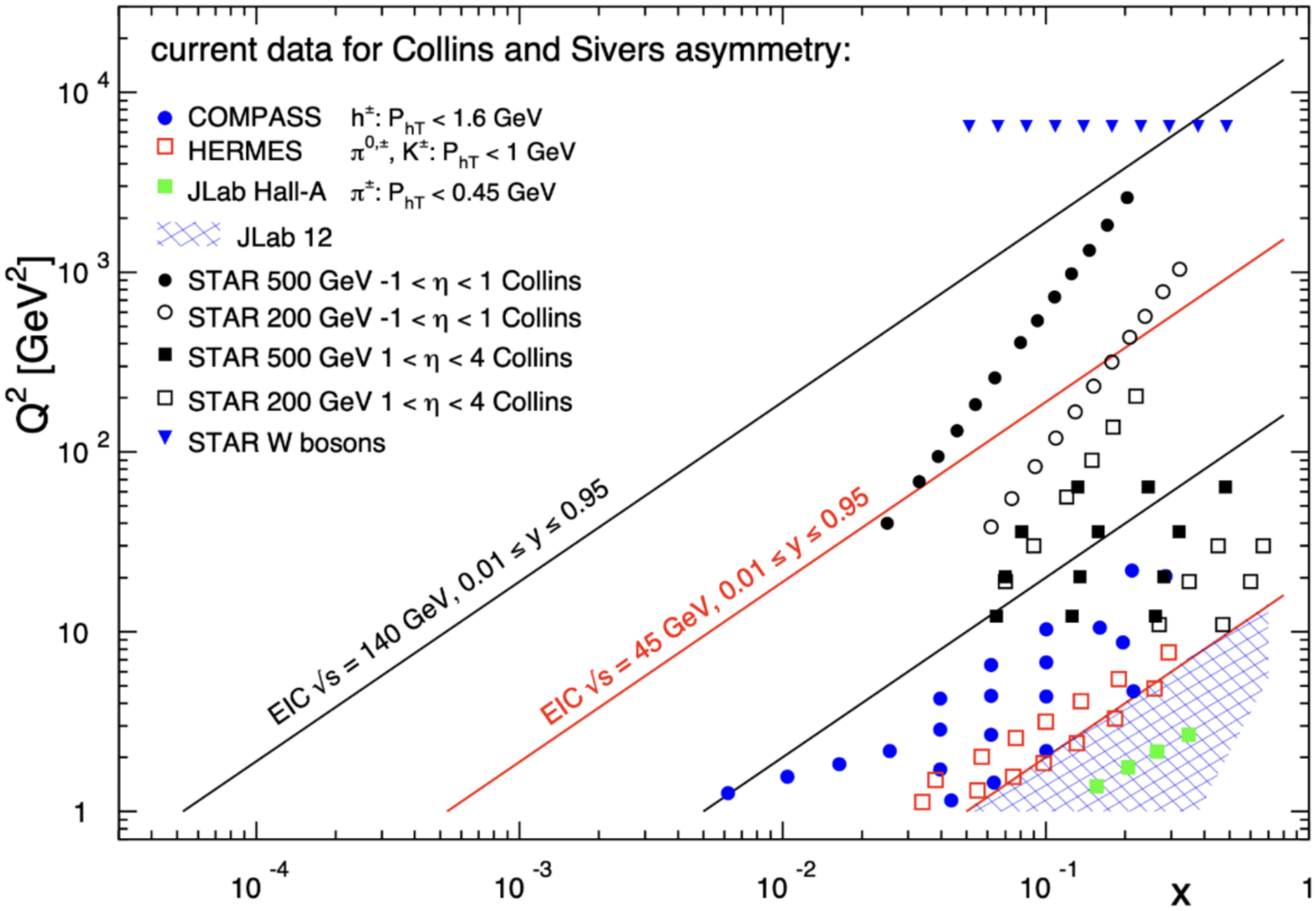}
\caption{ Left: The impact of RHIC data to constrain gluon helicity \cite{DSSV:2014,DSSV:prel}.
Right: The $x$-$Q^{2}$ probed with data from the future
EIC and Jlab-12 GeV as well as the current semi-inclusive deep inelastic scattering (SIDIS)
data and the jet and $W$-boson data from RHIC. All data are
sensitive to the Sivers function and transversity
times the Collins fragmentation function (FF) in the TMD formalism.}
\label{fig:deltaGimpact}
\end{figure*}

RHIC has driven the exploration of the fundamental structure of strongly interacting matter into new territory and will continue to enable advances in the field for years to come. These explorations have always thrived on the complementary nature of lepton scattering and purely hadronic probes. This is demonstrated clearly in the flagship measurements of the gluon and sea-quark helicity distributions that are discussed in detail in Section~\ref{CollinearChapter}. The sea-quark program exploited the advantages afforded by high energy hadron beams, using $W^{+/-}$ production to reveal the flavor asymmetry of the $\Delta\bar{u}$ and $\Delta\bar{d}$ distributions without the complications of fragmentation effects. Similarly, reconstructed jet and pion asymmetries were used for the first time to directly probe gluon interactions in proton-proton collisions, discovering a sizable gluon helicity distribution in the region $x > 0.05$, as shown in the left panel of Fig.\@ \ref{fig:deltaGimpact}.  

The RHIC Cold QCD program leverages the techniques and tools developed in the high profile helicity program to open new frontiers in the rapidly evolving field of transverse spin physics. For example, the reconstruction of $W$ bosons in transversely polarized proton collisions is used to test the predicted sign change of the Sivers' function and to provide the first constraints on the sea-quark Sivers functions. Hadron-in-jet asymmetries, measured for the first time at RHIC, and di-hadron asymmetries provide access to the collinear quark transversity distributions, as well as the transverse momentum dependent (TMD) Collins Fragmentation Function (hadron-in-jet) and collinear Interference Fragmentation Functions (di-hadron) in the final state. These new channels, and many more, are discussed in detail in Section~\ref{3DImageChapter}. Again, the transverse spin program exploits the complementarity of the high energy hadron collider configuration by accessing distributions originally measured in lepton scattering experiments, but in a different kinematic regime, allowing for new insights into universality, factorization and TMD evolution. 

%\begin{multicols}{2}
%\end{multicols}

As the realization of a future EIC draws closer, there is a growing scientific imperative to complete a set of ``must-do'' measurements in $pp$ and $pA$ collisions in the remaining RHIC runs. The ongoing RHIC Cold QCD program will build on the accelerator's unique ability to collide a variety of ion beams in addition to polarized protons, and a detector with wide kinematic coverage that has been further enhanced through an upgrade at forward rapidities consisting of electromagnetic and hadronic calorimetry as well as tracking.  The new detectors will enable $e/h$ discrimination with charge-sign determination and full jet reconstruction in the forward direction for the first time, allowing RHIC to extend the full complement of the existing transverse spin program into new kinematic regimes!  This will expand the existing transverse spin program into both lower and higher $x$ domains, as illustrated in the right panel of Fig.\@ \ref{fig:deltaGimpact}. In addition to the expanded transverse spin program, RHIC will be able to further explore exciting new signatures of gluon saturation and non-linear gluon dynamics (see Section~\ref{CollinearChapter}). The ratios of forward Drell-Yan and photon-jet yields in $pp$ and $pA/AA$ collisions are clean probes of nuclear modifications to initial state parton distributions as well as gluon saturation effects. All of these measurements rely critically on the successful completion of scheduled RHIC operations before the shutdown in 2025.

While the remaining RHIC Cold QCD program is unique and offers discovery potential on its own, it is also essential to fully realize the scientific promise of the EIC. These data will provide a comprehensive set of measurements in hadronic collisions that, when combined with EIC data, will establish the validity and limits of factorization and universality. The separation between the intrinsic properties of hadrons and interaction dependent dynamics, formalized by the concept of factorization, is a cornerstone of QCD and largely responsible for the predictive power of the theory in many contexts.  While this concept and the associated notion of universality of the quantities that describe hadron structure has been successfully tested for unpolarized and - to a lesser extent - longitudinally polarized parton densities, its experimental validation remains an unfinished task for much of what the EIC is designed to study, namely the three-dimensional structure of the proton and the physics of dense partonic systems in heavy nuclei. To establish the validity and the limits of factorization and universality, it is essential to have data from both lepton-ion and proton-ion collisions, with an experimental accuracy that makes quantitative comparisons meaningful. The final experimental accuracy achieved with the data collected during this final RHIC campaign will enable quantitative tests of process dependence, factorization and universality by comparing lepton-proton with proton-proton collisions. When combined with data from the EIC, it will provide a broad foundation to a deeper understanding of Quantum Chromodynamics.

\newpage
\subsection{Recommendations and Initiatives}

 The RHIC Cold QCD community proposes the following recommendations. These proposals were presented, discussed and received strong support at the QCD Town Hall Meeting in September of 2022. \\

\begin{enumerate}
    \item Continued  funding of RHIC operations to enable collection of the last $pp$, $pA$ and $AA$ datasets that are required for completion of the RHIC Hot and Cold QCD missions.
    \item Continued strong funding of the RHIC Hot and Cold QCD experimental analysis groups well beyond the final operation of RHIC. This will enable effective and timely analysis and publication of the wealth of data collected, and to be collected. The unprecedented originality of the RHIC data sets have discovery potential on their own and are critical to fully accomplish the scientific mission of the EIC. Continued support of the experimental groups is also critical to ensure the continued effective training of the next generation of experimentalists in preparation for EIC operations.
    \item Continued strong funding of the RHIC Hot and Cold QCD theoretical groups and collaborations well beyond the final RHIC operations to ensure that the knowledge generated by analyses of the RHIC data are fully incorporated into the next-generation of theoretical interpretations.
\end{enumerate}

%% file: HadronBeams.tex
\label{HadronBeamsChapter}
%\begin{multicols}{2}
The Relativistic Heavy Ion Collider, RHIC, has the unique capability to collide polarized protons at center-of-mass energies up to 510 GeV.
High beam polarizations are an important prerequisite for the efficient and timely execution of the physics program at RHIC, since the figure of merit for any spin-dependent observable is directly proportional to the square of the polarization of the beam and the number of events.
For double-spin observables, it is the product of the two beam polarizations which enters into the figure of merit, so losses that affect both beams similarly are potentially even more severe.

The BNL collider complex is using a set of accelerators between the proton (ion) source and the main RHIC collider.
The proton beams are injected in up to 120 bunches which can collide in several different interaction points.
Proton-proton collisions can reach center-of-mass energies up to 510 GeV.
Other collision energies in the past have included 64 and 200 GeV with longitudinal or transverse beam polarizations.
Proton bunch intensities reach $2.7\cdot10^{11}$, resulting in peak luminosities of about $5\cdot10^{32}~\mathrm{cm}^{-2}\mathrm{s}^{-1}$ with average beam polarizations of $\left<P\right>\approx 55\%$.
The recent RHIC Run-22 delivered about $800~\mathrm{pb}^{-1}$ with an average $68~\mathrm{pb}^{-1}$ per week (the luminosity was limited to $1.27\cdot10^{32}~\mathrm{cm}^{-2}\mathrm{s}^{-1}$ by request from the experiment).

\subsection{Preparation and Preservation of Hadron Beam Polarization}

Unlike electron beams, hadron beams at the RHIC energy scale lack any significant synchrotron radiation mechanism for self-polarizing or natural emittance damping.
This means that the production of high-energy, high-brightness, highly polarized beams consists largely of creating intense, low emittance polarized beams at the source and carefully preserving the beam quality through all stages of acceleration.
Developments in source technology and accelerator physics are therefore both of key importance in supporting the RHIC spin physics program.

The RHIC polarized proton beams are created in an optically pumped polarized ion source (OPPIS \cite{OPPIS}).
The source has undergone continuous development during the RHIC era, including a major upgrade in 2013.
Source technology upgrades include changing from an electron cyclotron resonance source to a fast atomic beam source (ABS), addition of a superconducting solenoid for enhanced polarization preservation, and installation of a novel pulsed electromagnetic vacuum valve.
The output beam of the source is actually an $H^{-}$ beam where the protons have been polarized. The upgraded source reliably produces beams of up to $10^{12}$ particles in a 300 $\mu$s pulse.
Proton polarizations of 82-84\% have been achieved out of the source as measured by a high precision carbon target polarimeter at the end of the 200 MeV Linac.
The source can change the spin orientation (vertically up and down) between consecutive pulses, allowing for an arbitrary pattern of spin directions along the RHIC bunch train.
The high pulse intensity \footnote{The pulse intensity is a factor of 3 higher than what is required in RHIC.} enables various {\it scraping} schemes in the downstream accelerators which reduce the intensity by scraping away high amplitude particles (transversely and longitudinally) but preserve a high brightness core with an intensity and emittance optimized for RHIC operations.

After creation in the source and acceleration in the 200 MeV Linac (to a total energy of 1.1 GeV), the $H^{-}$-beam is injected into the Booster synchrotron via a charge-exchange injection process which strips the electrons with a carbon foil and produces the proton beam.
From there the beam is accelerated in three successive synchrotrons.
First the protons are accelerated in the Booster to 2.3 GeV, then in the Alternating Gradient Synchrotron (AGS) to 23 GeV and finally in the RHIC rings themselves to a top energy of up to 255 GeV. The magnetic fields that produce the steering and focusing necessary to confine the beam during acceleration will also induce motion of the spin vectors of the particles as they precess about the local field lines.
This combination of particle motion and spin motion in a periodic system like a synchroton ring introduces resonance conditions which can cause rapid polarization losses.
Resonances occur whenever the natural frequency of the spin motion (characterized as the number of spin precessions per turn, or spin tune) is equal to some frequency with which the spin perturbing magnetic fields are sampled.
A magnetic field that deflects a proton by an angle $\theta$ will precess the spin vector by an angle $G\gamma\theta$, where $G$ is the anomalous gyromagnetic ratio ($G=1.79$ for protons) and $\gamma$ is the relativistic factor.
The spin motion in magnetic fields is therefore highly energy dependent and during acceleration to top energy in RHIC these resonance conditions are met very often.

In order to reduce polarization losses, the spin motion of particles can be manipulated such that some resonances are completely avoided.
In RHIC this is accomplished with helical dipoles (so-called Siberian snakes).
Two helical dipoles magnets, located diametrically opposite one another, each provide a full spin-flip (from up to down or vice versa) in a single passage (for each of the two rings).
The result is that the spin tune is made energy-independent (fixed at 1/2) and there is complete cancellation of all first order depolarizing resonance terms \cite{RHIC_config_pp}.
Even in the presence of full snakes, however, depolarization is still possible due to higher order resonances called {\sl snake resonances}.
Minimizing the effects of these resonances requires careful control of the betatron tune (the natural frequency of the transverse particle oscillations) during acceleration.
In RHIC this is done using a fast phase-locked loop tune feedback system, in operations since 2006 \cite{TuneFeedback} which enables acceleration very close to a low-order betatron resonance without loss of intensity or emittance dilution, and which maximizes the distance to the nearest depolarizing snake resonance.
This is an excellent example of developments in accelerator physics and beam control being driven by the stringent demands of a polarized collider physics program.

In the AGS, resonance avoidance is somewhat more complicated because there is not sufficient space in the magnetic lattice for snakes that provide a full spin-flip.
Instead, the AGS uses two weaker {\sl partial} snakes, which rotate the spin through an angle less than $180^\circ$ ($18^\circ$ and $10^\circ$, respectively).
In this case the spin tune is still energy dependent, but is prevented from taking a small range of values \cite{Huang_DualSnakes}.
By careful control of the betatron tune, this allows the strongest resonance to be avoided. A fast tune jump minimizes the polarization loss from the many weak residual resonances that are driven by the partial snakes themselves.
The Booster, by contrast, only has two resonance crossings, which are each handled by an orbit harmonic correction scheme, rather than specialized snake magnets.

The strongest resonances encountered during acceleration are dependent on the beam emittance, with the larger amplitude particles experiencing greater depolarization.
This makes emittance control for a polarized beam particularly important since it affects two components of the collider performance: luminosity and polarization. 
A variety of techniques are employed in the RHIC accelerator complex to prevent emittance growth.
Bunch lengthening manipulations, like dual harmonic RF schemes (in AGS and Booster) and the addition of the low frequency 9 MHz RF system in RHIC lower the peak current of the bunches and prevent emittance increases due to space charge forces and electron cloud buildup.
A strong driver of emittance growth during RHIC polarized proton stores is the nonlinear defocusing force of each beam upon the other (the beam-beam effect) at the point of collision, which can cause fast emittance blowup immediately upon steering the beams into one another. Starting in Run-15, this fast blowup was suppressed by a pair of electron lenses, which provide an equal but opposite nonlinear force \cite{ELens}.

In addition to preserving the beam polarization, the spin physics program requires specialized instrumentation and methods to measure aspects of the spin dynamics like the spin tune and the spin direction at key locations in the ring.
RHIC is the first polarized beam facility to measure the spin tune non-destructively using coherent excitation of the spin motion.
An interleaved sequence of 5 RF dipoles and 4 DC dipoles produces a driven coherent precession of the particle spin vectors.
The pC polarimeters deliver a measurement of the asymmetry synchronized to the phase of the driving field.
The spin tune is then calculated from the relationship between the driven and measured amplitudes.
Since this is an adiabatic excitation, it can be reversed without loss of polarization, which allows scans and repeated measurements at top energy without costly refill times \cite{SpinTuneMeter}.
This integration of accelerator physics methods with polarimetry allowed optimization of the accelerator lattice to optimize the spin tune and its spread.

These efforts will continue to be important in the EIC era, and a host of further developments will continue through the end of the RHIC program.
The hadron spin dynamics and equipment in the EIC are more complicated than those of RHIC. %Additional snakes in RHIC will improve the resonance structure, but place additional requirements on the magnetic lattice.
The altered IP geometry produces additional complications for the spin rotation manipulation, including change of spin direction in the arcs, which will have to be measured, controlled and verified.
A more complete partial snake resonance compensation for the AGS is planned for commissioning in Run-24.
Furthermore, the production and acceleration of polarized Helium-3 requires extensive source development and use of the full toolkit of resonance avoidance including re-introduction of methods previously used for polarized protons, like use of an AC dipole resonance crossing, currently under development in the Booster.

\subsection{Polarimetry of High-Energy Hadron Beams}

The experiments at RHIC require that the beam polarizations are known with good accuracy.
Double helicity asymmetries are typically small ($O(10^{-3})$) and knowledge of the relative luminosity is the leading uncertainty for these observables.
Early estimates have set a requirement of less than $\sigma(P)/P=4\%$ for the relative polarization uncertainty.
With increased interest in large transverse single-spin asymmetries and high luminosity data sets, the achieved benchmark has been lowered to $1.4\%$ in recent years \cite{RHIC_BeamPolarizations_2018}.

The beam polarizations at RHIC are measured through a combination of absolute and fast polarimeters.
Both are based on the detection of recoil particles from elastic scattering at low energies, where a spin-dependent asymmetry is introduced through a spin-flip in the Coulomb-Nuclear interference region.
The determination of the absolute beam polarization makes use of a polarized atomic hydrogen gas jet target, HJET, resulting in an uncertainty of a $3-4\%$ over the course a typical RHIC fill (8 hours long).
The target polarization is prepared from a state-of-art atomic beam source in a sextupole magnet system in combination with an RF-transition unit, which optimizes the atomic target density and polarization, $P\approx 96\%$.
The remaining molecular fraction of $H_2$ in the target has been the major source of uncertainty in the determination of the beam polarization.
The recent significant improvement is a major step towards the applicability of the existing system at the EIC, where a polarization uncertainty of 1\% or better is required.

The HJET is complemented by fast measurements with Carbon fiber targets every few hours which allow for the tracking of polarization decay from injection to the end of each RHIC fill.
The time-dependent knowledge of the beam polarization is important for the correct use in the experiments, where the luminosity-weighted polarization can be significantly different from the time-average value.

The ultra-thin Carbon targets scan transversely through the beam bunches during each set of measurements, thereby providing a picture of the transverse beam polarization profile itself.
While it was originally assumed that the transverse polarization profile is flat, the measurements show that the polarization indeed peaks in the center of the bunches, in both the horizontal and vertical directions.
This information, again, is essential for the proper determination of the beam polarizations in collision at the experiments.
The convolution of bunch intensities with the polarization profile results in a higher polarization value than that measured with the HJET.
Maybe even more surprising, a longitudinal polarization profile of the proton bunches has been measured with the HJET and the Carbon polarimeters independently.
The longitudinal polarization dependence is smaller than the transverse profile; in addition, the polarization is lower towards the center of the bunch.

Polarimetry at RHIC relies on a good understanding of the spin dynamics and the stable spin direction of the accelerator.
The transverse direction of the polarization vector (with respect to the beam momentum) at the collision points in the experiments is confirmed and monitored through local polarimetry.
The concept is based on neutron production at very forward directions, measured in the Zero Degree Calorimeters.
This method was discovered in the first polarized proton collisions at RHIC, and it was essential for the successful commissioning of the spin rotators for longitudinally polarized experiments.

Control and verification of the spin direction is important at the experimental collision points as well as at the location of the polarimeters.
In a lattice like RHIC with full snakes, the design stable spin direction (the direction about which misaligned spin vectors will precess from turn to turn), is vertical.
Pairs of additional helical dipoles, called spin rotators, that flank the interaction points of the large experiments in the RHIC ring can rotate the spin locally, providing longitudinally or radially directed polarization at those specific locations, while leaving the polarization in the rest of the ring unperturbed.
The spin direction can deviate from the intended orientations for a number of reasons, including errors in the helical dipole fields and misalignment of quadrupoles in the lattice.
Over the years, systematic scans of the helical dipole settings and the beam energy have helped to understand and characterize the source of these deviations.
These efforts were particularly vital in shaping the response to a failure of two of the four magnets that make up one of the snakes in the Blue ring in Run-22.
Meeting the physics requirements with only the remaining two magnets required compensation with the other functioning snake, a change in the store energy to minimize the resulting (and now strongly energy dependent) deviation of the spin direction, and development of new methods of spin direction measurement.
Since the local polarimetry at STAR and the pC polarimeters are only sensitive to the transverse components of the stable spin direcion, measurements were developed using the spin rotators (at STAR) and a horizontal orbit angle (at the pC) to rotate the hidden longitudinal direction into the transverse plane \cite{RHIC_Run22}.

Proton polarimetry at RHIC does not only provide vital input for the experiments and fast feedback to the collider during beam development and regular operations.
It has also delivered surprising results, which further our general understanding of spin-dynamics in particle accelerators and storage rings.
The intricate correlation of the spin tune, stable spin-direction, and the polarization lifetime has been studied over many years, but it proved to be of special importance during RHIC Run-22 when one of the Siberian snakes had a partial failure.
Based on past experiences and through ingenious combination of the remaining snake parts with spin rotators and the polarimeters, it was possible to determine a setup that showed no significant loss of beam polarization at flattop energy.
All of this information will directly benefit the future Electron-Ion Collider and any other polarized accelerator for medical or other applications.

%\end{multicols}

%% file: Collinear.tex
\begin{itemize}
\label{CollinearChapter}

\item RHIC high precision longitudinally polarized proton-proton data for a variety of probes and center of mass energies have played a decisive role in constraining the sea-quark and gluon helicity distributions in the proton.
\item $W$ production in longitudinally polarized $pp$ collisions revealed the existence of a flavor asymmetry in the polarization in the sea of light anti-quarks with $\Delta\bar{u}$ being positive, while $\Delta\bar{d}$ is negative. 
\item Collisions at center of mass energies of 200 GeV provided the first evidence that the gluons inside a proton are polarized. Data from the RHIC run in 2009, when included in global analysis, showed that gluons carry approximately 40\% of the proton spin in the region where the gluon carries more than 5\% of the proton momentum ($x > 0.05$) at $Q^2=10\,\mathrm{GeV}^2$.
%\cite{DSSV:2014, NNPDF:2014}.
\item The published and preliminary results based on data collected in 2012, 2013 and 2015 at center of mass energies of 200 and 510 GeV reduce the present uncertainties on gluon helicity $\Delta g$ even further, providing more insights in the region of momentum fraction $x$ between about 0.01 to 0.5 of the momentum of a polarized proton. 
\item {STAR $pp$ and $pA$ forward di-hadron correlation results pioneered the observation of nonlinear gluon dynamics dependence on the nuclear mass number $A$.
%\cite{STAR:2021fgw}.
Higher-precision measurements will be performed with the STAR forward upgrade to further explore nonlinear gluon dynamics. All the studies provide the baseline for searching for gluon saturation at the future EIC.}
\end{itemize}

\newpage
%\begin{multicols}{2}
\subsection{$W$ $A_L$ and sea quark polarization}

The STAR and PHENIX Collaborations have concluded the measurements of the parity-violating spin asymmetry in the production of weak bosons from collisions with one of the proton beams polarized longitudinally \cite{STAR:2010xwx, PHENIX:2010rkr, STAR:2014afm, PHENIX:2015ade, PHENIX:2018wuz, STAR:2018fty}. In 510 GeV center-of-mass proton-proton collisions at RHIC, $W^{+}$ bosons are produced primarily in the interactions of $u$ quarks and $\bar{d}$ antiquarks, whereas $W^{-}$ bosons originate from $d$ quarks and $\bar{u}$ antiquarks. The longitudinal single spin-asymmetry ($A_{L}$) measurements of the decay positrons provide sensitivity to the $u$ quark and $\bar{d}$ helicities in the proton, whereas the decay electrons do so for the $d$ and $\bar{u}$ helicities. Combined, they make it possible to delineate the light quark and antiquark polarizations in the proton by flavor.

%\begin{figure}[H]
\begin{SCfigure}[][h]
    \centering
    \includegraphics[width=0.5\columnwidth]{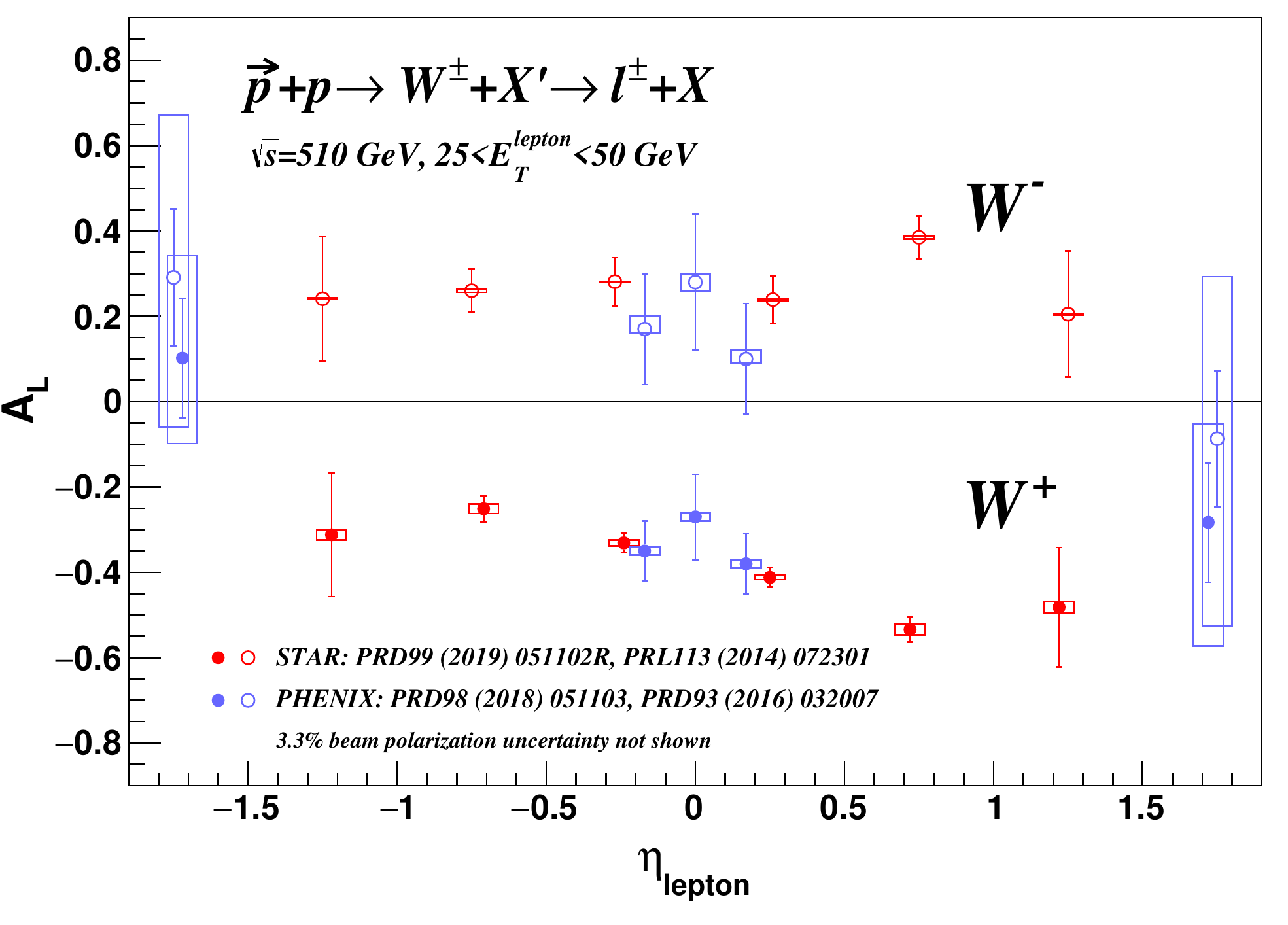}
    \caption{Longitudinal single-spin asymmetries, $A_{L}$, for $W$
production as a function of the lepton pseudorapidity, $\eta_{\mathrm{lepton}}$, for the combined STAR and PHENIX data
samples \cite{STAR:2014afm,PHENIX:2015ade,PHENIX:2018wuz,STAR:2018fty}.}
    \label{fig:WAL_RHICall}
%\end{figure}
\end{SCfigure}

%\begin{figure}[H]
\begin{SCfigure}[][h]
    \centering
    \includegraphics[width=0.5\columnwidth]{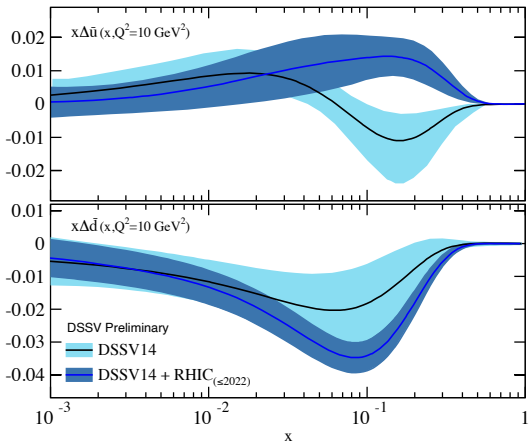}
    \caption{The impact of the RHIC $W$ $A_L$ results on $\bar{u}$ (top) and $\bar{d}$ (bottom) polarizations as a function of $x$ at a scale of $Q^{2}$= 10 GeV$^{2}$. The black curves with the $1\sigma$ uncertainty bands marked in light blue show the results from the DSSV14 global fit \cite{deFlorian:2014yva} and the blue curves with $1\sigma$ uncertainty bands in dark blue show the results for the new preliminary DSSV fit \cite{DSSV:prel} including the RHIC $W$ data~\cite{STAR:2014afm, PHENIX:2015ade, STAR:2018fty}.}\label{fig:W_impact_ud}
%\end{figure}
\end{SCfigure}

\begin{SCfigure}[][h]
%\begin{figure}[H]
    \centering
    \hspace{-0.5cm}
    \includegraphics[width=0.5\columnwidth]{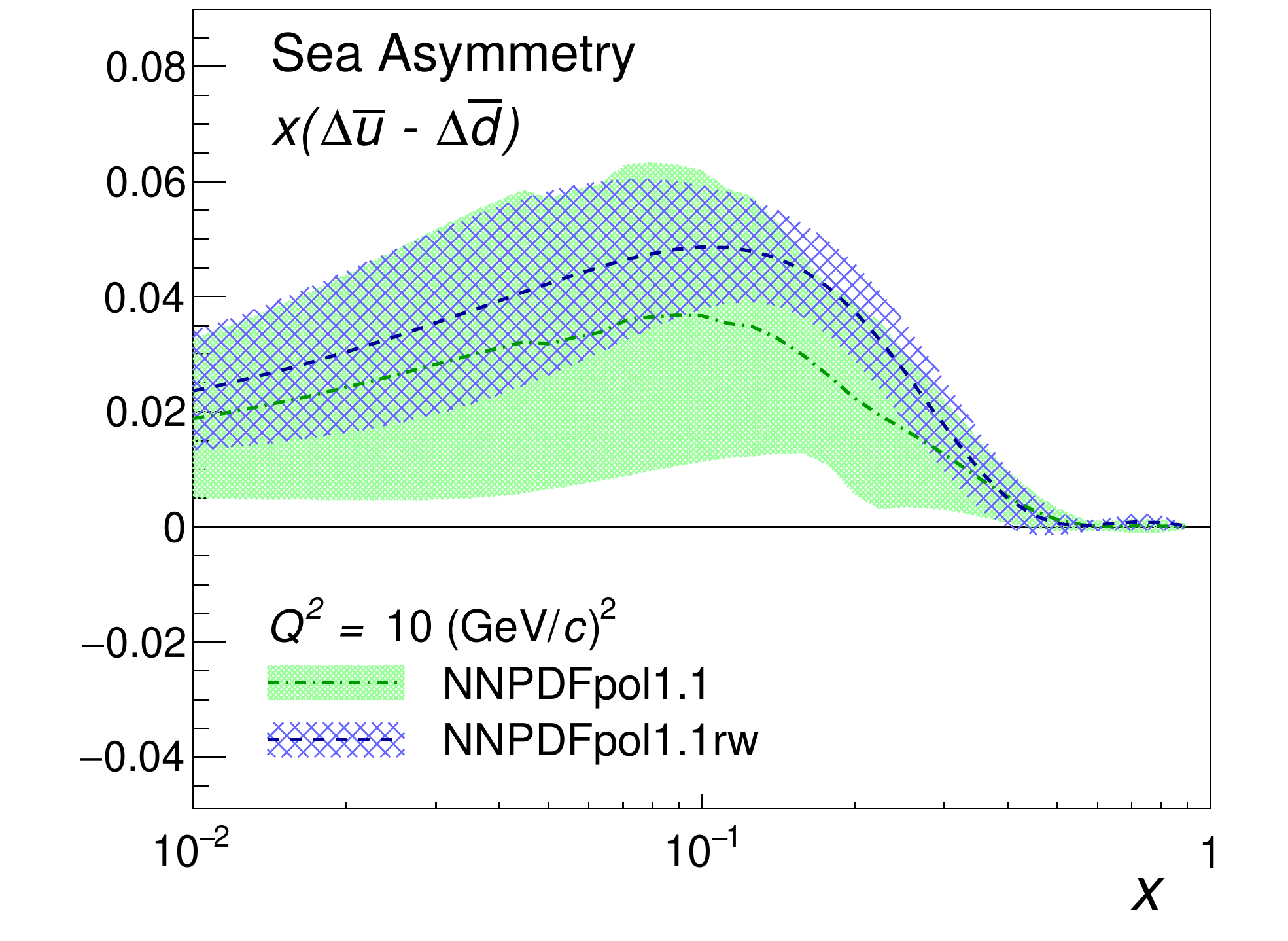}
    \caption{The difference of $\bar{u}$ and $\bar{d}$ polarizations as a function of $x$ at a scale of $Q^{2}$= 10 GeV$^{2}$ before and after NNPDFpol1.1 \cite{Nocera:2014gqa} reweighting with STAR 2013 $W$ $A_{L}$~\cite{STAR:2018fty}. The green band shows the NNPDFpol1.1 results \cite{Nocera:2014gqa} and the blue hatched band shows the corresponding distribution after the STAR 2013 $W$ data are included by reweighting. }\label{fig:W_impact_sea}
\end{SCfigure}

These measurements shed light on understanding of the light quark polarizations $-$ one of the two initial motivations for the spin-physics program at RHIC. The data, shown in Fig. \ref{fig:WAL_RHICall}, are the final results from STAR and PHENIX on this topic \cite{STAR:2014afm, PHENIX:2015ade, PHENIX:2018wuz, STAR:2018fty} that combine all the published data obtained in 2011, 2012, and 2013. The impact of the RHIC $W$ data on the sea quark helicity distributions $\Delta\bar{u}$ and $\Delta\bar{d}$ is presented in Fig.~\ref{fig:W_impact_ud}. The plot shows the impact of the RHIC $W$ data~\cite{STAR:2014afm, PHENIX:2015ade, STAR:2018fty} from the new global fit by the DSSV group including also the recent jet, dijet, and pion data \cite{Adam:2018pns, Adam:2019aml, JetsALL:2015, STAR:2021mqa, starEndcapDijetALL2013, Adare:2015ozj, PHENIX:2020trf} (that constrain mostly the gluon helicity). The sea quark $\bar{u}$ helicity $\Delta\bar{u}$ is now known to be positive and $\Delta\bar{d}$ is negative. The STAR 2013 data \cite{STAR:2018fty} were also used in the reweighting procedure with the publicly available NNPDFpol1.1 PDFs \cite{Nocera:2014gqa}. The results from this reweighting, taking into account the total uncertainties of the STAR 2013 data and their correlations, are shown in Fig. \ref{fig:W_impact_sea} as the blue hatched bands. The NNPDFpol1.1 uncertainties are shown as the green bands for comparison. As seen from the plot, the data have now reached a level of precision that makes it possible, for the first time, to conclude that there is a clear asymmetry between the helicity distribution of $\bar{u}$ and $\bar{d}$, and it has the opposite sign from the $\bar{d}/\bar{u}$ flavor asymmetry in the unpolarized sea.

\begin{comment}
\begin{figure}[H]
    \centering
    \includegraphics[width=\columnwidth]{fig/ppW.png}
    \caption{Longitudinal single-spin asymmetries, $A_{L}$, for $W$ production for the combined STAR and PHENIX data~\cite{STAR:2014afm, PHENIX:2015ade, STAR:2018fty}. The DSSV14 evaluations \cite{deFlorian:2014yva} are marked with light blue and the dark blue bands show the results for the new DSSV fit including the RHIC W data~\cite{STAR:2014afm, PHENIX:2015ade, STAR:2018fty}. Red curves show the JAM solution for the negative gluon polarization $\Delta g<0$ \cite{Zhou:2022wzm} calculated by the DSSV group (see discussion in Section \ref{sec:gluon-helicity}).}
    \label{fig:WAL_RHICimpact}
\end{figure}

The RHIC $W$ $A_{L}$ data not only help to constrain the sea quark helicity, but also strongly disfavor the large and negative gluon polarization solution suggested in the recent JAM collaboration \cite{Zhou:2022wzm} fit. 
Figure. \ref{fig:WAL_RHICimpact} shows the data are in good agreement with preliminary fit results from DSSV group based on positive gluon polarization, meanwhile, they are in disagreement with the evaluations based on negative gluon polarization from JAM collaboration \cite{Zhou:2022wzm}. See discussion in Section \ref{sec:gluon-helicity}.
\end{comment}
\subsection{Double helicity asymmetries $A_{LL}$ and gluon polarization}
\label{sec:gluon-helicity}
The measurement of the gluon polarization inside protons has been a major emphasis of the longitudinally polarized RHIC program. At RHIC, gluon polarization can be accessed by measurements of the spin-dependent %rates of jets \cite{Abelev:2006uq,
rates of production of jets \cite{Abelev:2006uq,Abelev:2007vt, Adamczyk:2012qj, Adamczyk:2014ozi, Adam:2019aml, JetsALL:2015, STAR:2021mqa}, dijets \cite{Adamczyk:2016okk, Adam:2018pns, Adam:2019aml, JetsALL:2015, STAR:2021mqa}, $\pi^0\mathrm{s}$ and charged pions, \cite{Adare:2007dg ,Adare:2008aa, Adare:2008qb, Abelev:2009pb, Adamczyk:2013yvv, Adare:2014hsq, Adare:2015ozj, Adam:2018cto, PHENIX:2020trf}, and direct photons~\cite{PHENIX:2022lgn}. Data from the RHIC run in 2009 have for the first time shown that gluons inside a proton are polarized with a strong constraint from the jet data at a center-of-mass energy of $\sqrt{s} = 200\,\mathrm{GeV}$~\cite{deFlorian:2014yva, Nocera:2014gqa}. Perturbative QCD analyses~\cite{deFlorian:2014yva, Nocera:2014gqa} of the world data, including 2009 inclusive jet and $\pi^0$ results, at next-to-leading order (NLO) precision, suggest that gluon spins contribute $\simeq\!40\%$ to the spin of the proton for gluon fractional momenta $x > 0.05$ at a scale of $Q^2 = 10\,(\mathrm{GeV}/c)^2$. Results for dijet production provide a better determination of the functional form of $\Delta g(x)$, compared to inclusive observables, because of better constraints on the underlying kinematics \cite{deFlorian:2019zkl}.

\begin{SCfigure}[][h]
    \centering
    \includegraphics[width=0.5\columnwidth]{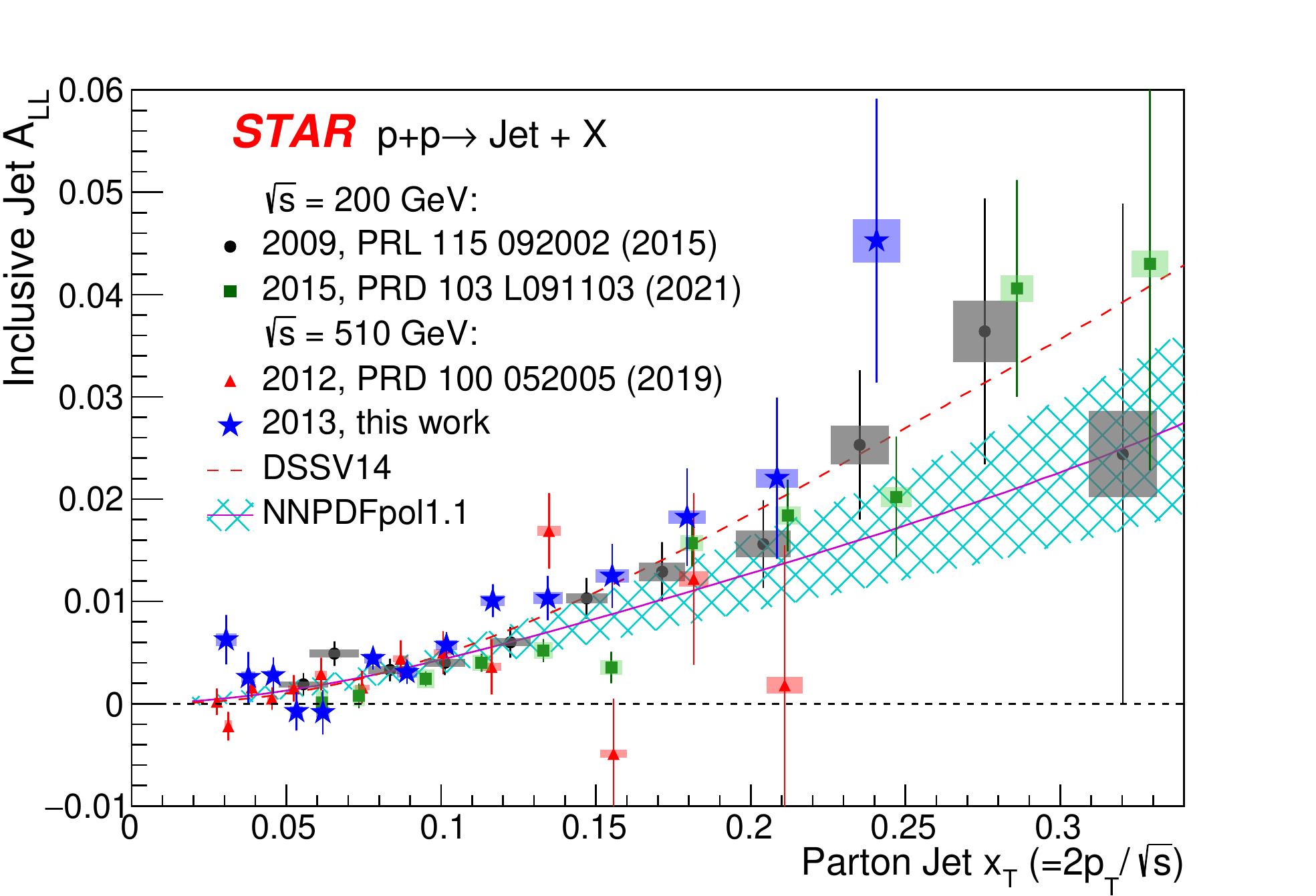}
    \caption{STAR results on inclusive jet $A_{LL}$ versus $x_{T}$ at $\sqrt{s}$ = 200 GeV \cite{Adamczyk:2014ozi, JetsALL:2015} and 510 GeV \cite{Adam:2019aml, STAR:2021mqa} at mid-rapidity from data collected in years 2009-2015, and evaluations from DSSV14 \cite{deFlorian:2014yva} and NNPDFpol1.1 (with its uncertainty) \cite{Nocera:2014gqa} global analyses. The vertical lines are statistical uncertainties. The boxes show the size of the estimated systematic uncertainties. Scale uncertainties from polarization (not shown) are $\pm$6.5\%, $\pm$6.6\%, $\pm$6.4\% and $\pm$6.1\% from 2009 to 2015, respectively. Source:~\cite{STAR:2021mqa}.}
    \label{fig:IncJet2013}
\end{SCfigure}

\begin{SCfigure}[][h]
    \centering
    %\hspace{-3cm}
    \includegraphics[width=0.5\columnwidth]{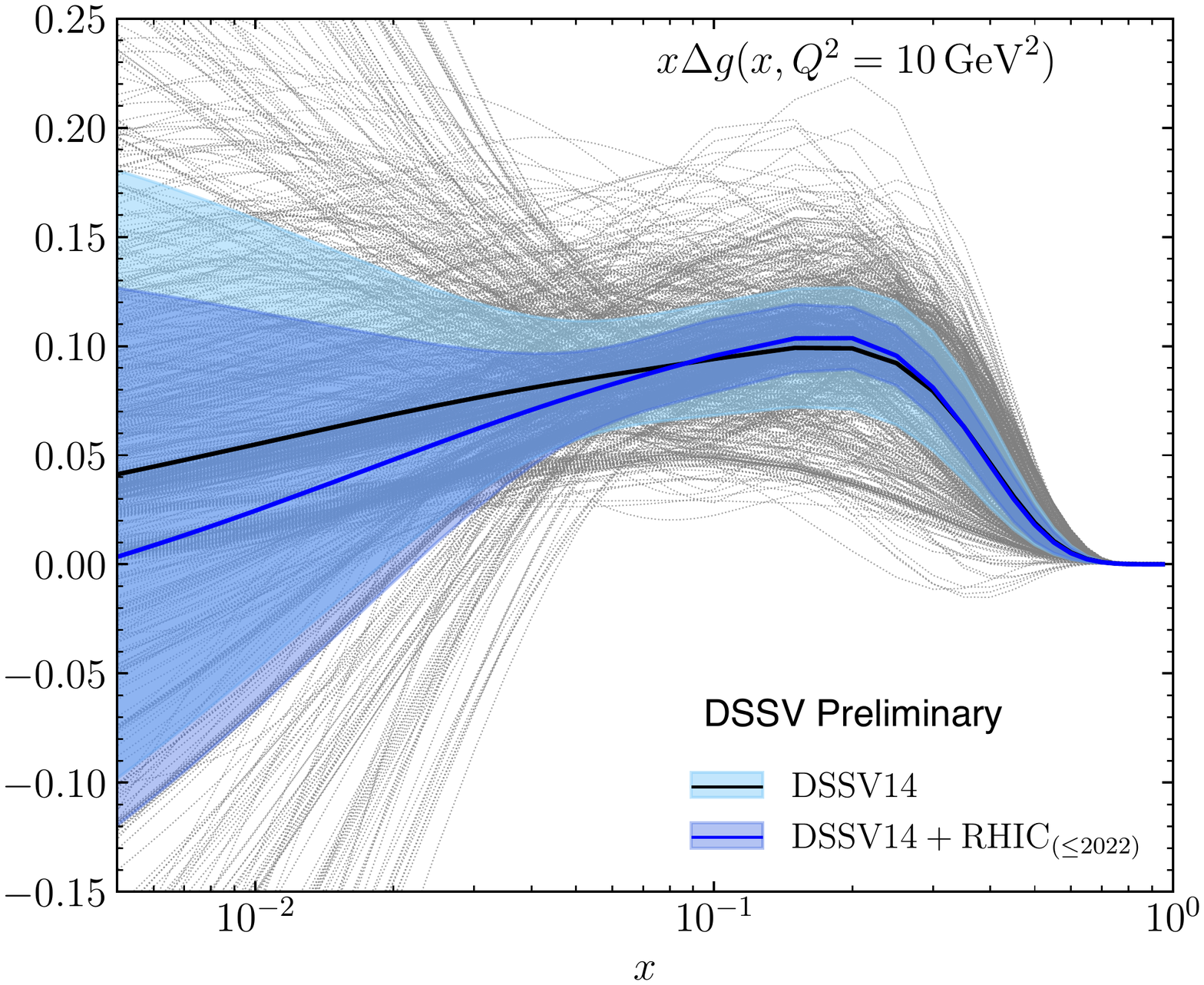}
    \caption{The impact of the recent jet and dijet\cite{Adam:2018pns, Adam:2019aml, JetsALL:2015, STAR:2021mqa, starEndcapDijetALL2013}, pion~\cite{Adare:2015ozj, PHENIX:2020trf} and $W$~\cite{STAR:2014afm, PHENIX:2015ade, STAR:2018fty} data on the $x$-dependence of the gluon helicity distribution at $Q^2=10\,\mathrm{GeV}^2$ based on the global fit by the DSSV group. The black curve with the $1\sigma$ uncertainty light blue band illustrates the DSSV14 results \cite{deFlorian:2014yva}, while the blue curve with $1\sigma$ uncertainty band in dark blue \cite{DSSV:prel} shows the preliminary results after the inclusion of the new data.}
    \label{fig:Gluon-impact}
\end{SCfigure}

Recent STAR results \cite{Adam:2019aml, JetsALL:2015, STAR:2021mqa} and preliminary results \cite{starEndcapDijetALL2012, starEndcapDijetALL2013} on longitudinal double-spin asymmetries of inclusive jet and dijet production at center-of-mass energies of 200 GeV (Run-15) and 510 GeV (Run-12 and Run-13) at mid and intermediate rapidity complement and improve the precision of previous STAR measurements. Figure \ref{fig:IncJet2013} shows recent STAR results on inclusive jet $A_{LL}$ versus $x_{T} = 2 p_T/\sqrt{s}$ at $\sqrt{s}$ = 200 GeV and 510 GeV at mid-rapidity from data collected in years 2009-2015, and evaluations from the DSSV14 \cite{deFlorian:2014yva} and NNPDFpol1.1 \cite{Nocera:2014gqa} global analyses. The overall impact of the recent jet and dijet~\cite{Adam:2018pns, Adam:2019aml, JetsALL:2015, STAR:2021mqa, starEndcapDijetALL2013}, pion~\cite{Adare:2015ozj, PHENIX:2020trf} and $W$~\cite{PHENIX:2015ade, STAR:2018fty} data on the $x$-dependence of the gluon helicity distribution at $Q^2=10\,\mathrm{GeV}^2$ based on the global fit by the DSSV group is presented in Fig.~\ref{fig:Gluon-impact}. The truncated moment of the gluon helicity from the new DSSV evaluations \cite{DSSV:prel} at $Q^2 = 10\,(\mathrm{GeV}/c)^2$ integrated with the range of $x \in (0.001,0.05)$ is $0.173(156)$ and in the range of $x \in (0.05,1)$ is $0.218(27)$ (at 68\% C.L.), which can be seen in the left panel of Fig. \ref{fig:deltaGimpact}.

\begin{SCfigure}[][h]
    \centering
    \includegraphics[width=0.5\columnwidth]{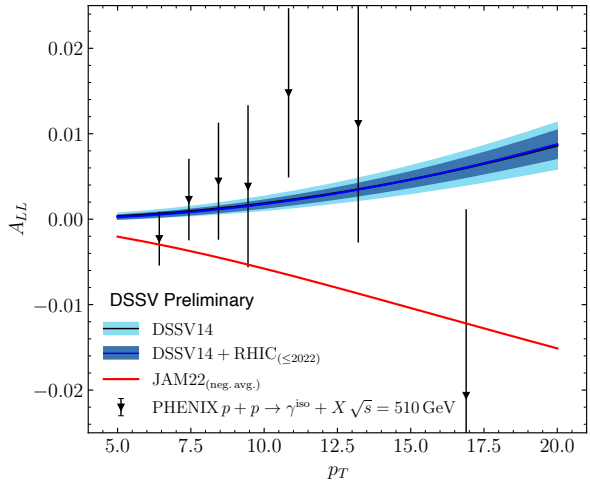}
    \caption{PHENIX double-helicity asymmetry $A_{LL}$ vs $p_T$ for isolated direct-photon production in polarized $pp$ collisions at $\sqrt{s}$=510 GeV at midrapidity ~\cite{PHENIX:2022lgn}. DSSV14 calculation is plotted as the black curve with the $1\sigma$ uncertainty band marked in light blue. The blue curve with $1\sigma$ uncertainty band in dark blue shows the impact of all the data sets included in the new preliminary DSSV fit \cite{DSSV:prel} as in Fig.~\ref{fig:Gluon-impact}. The curve for JAM $\Delta g<0$ solution \cite{Zhou:2022wzm} was calculated by W.Vogelsang.}
    \label{fig:PHENIX_PhotonALL}
\end{SCfigure}

\begin{SCfigure}[][h]
    \centering
    \includegraphics[width=0.5\columnwidth]{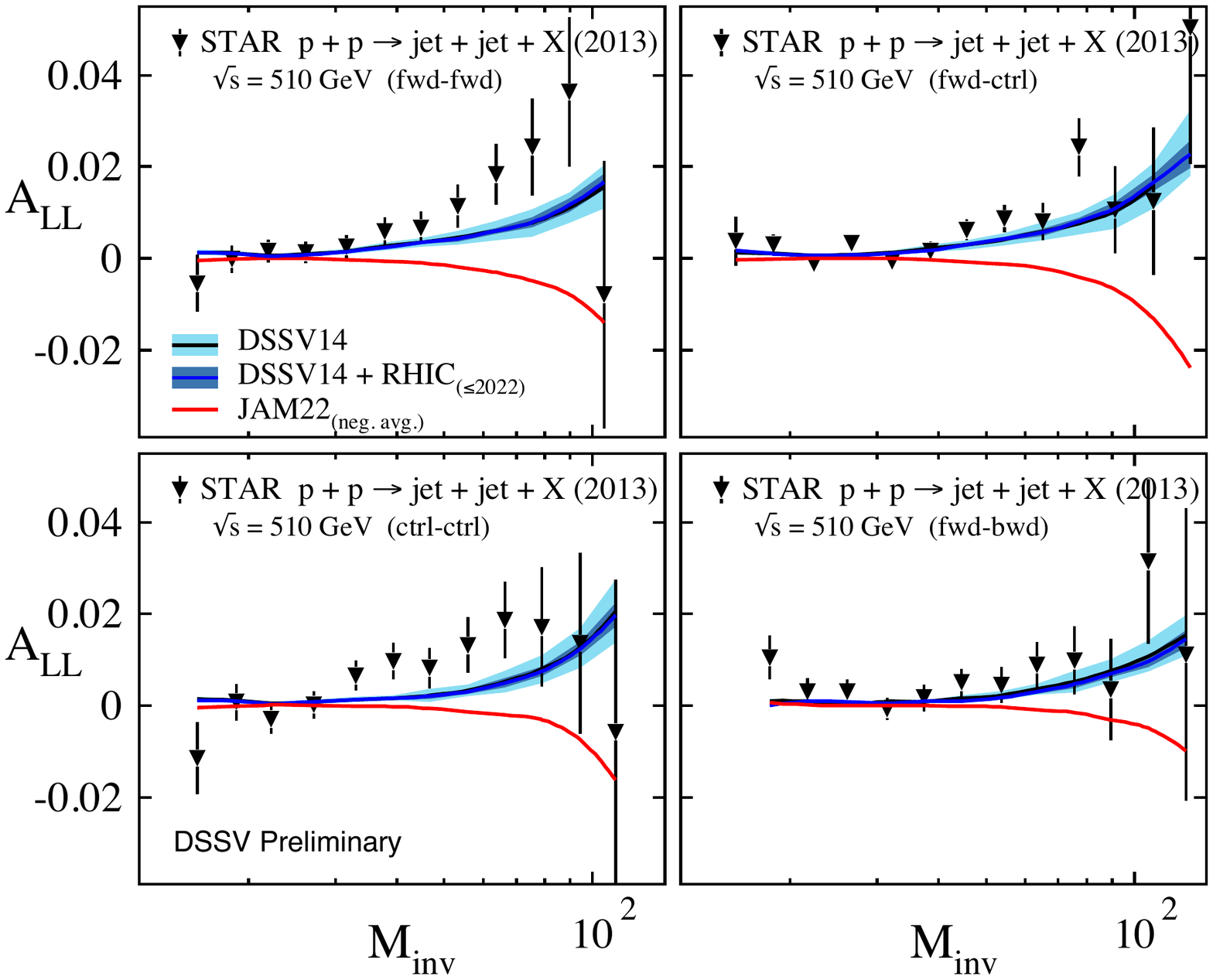}
    \caption{STAR double-helicity asymmetries $A_{LL}$ for dijet production vs dijet invariant mass $M_{inv}$ in polarized $pp$ collisions at $\sqrt{s}$=510 GeV at midrapidity from 2013 data set~\cite{STAR:2021mqa}. 
    DSSV14 evaluation~\cite{deFlorian:2014yva} is plotted as the black curve with the $1\sigma$ uncertainty band marked in light blue. The blue curve with $1\sigma$ uncertainty band in dark blue shows the impact of all the data sets included in the new preliminary DSSV fit \cite{DSSV:prel} as in Fig.~\ref{fig:Gluon-impact}. The red curves show the JAM $\Delta g<0$ solution \cite{Zhou:2022wzm} calculated by the DSSV group.}
    \label{fig:STAR_DijetALL}
\end{SCfigure}

The truncated moment of the gluon helicity integrated from $x = 0.0071$ to 1 at $Q^2 = 10\,(\mathrm{GeV}/c)^2$ from the recent JAM global QCD analysis \cite{Zhou:2022wzm} including a subset of RHIC data, i.e., STAR inclusive jet results, and assuming the SU(3) flavor symmetry and PDF positivity is $0.39(9)$. Authors of \cite{Zhou:2022wzm} also discuss the possibility of the solution with negative gluon contribution if the PDF positivity constraint is removed from the global fit. They argue that there is no fundamental theoretical requirement for PDF to be positive at all values of $x$, and therefore it would be highly desirable to have an observable which is linearly sensitive to gluon helicity distribution. Direct photons coming mainly from the quark-gluon Compton process and dijets narrowing down the parton kinematics are ideal probes to distinguish between positive and negative gluon helicity solutions. Figure~\ref{fig:PHENIX_PhotonALL} demonstrates the preference of positive solution with the PHENIX direct photon $A_{LL}$ data~\cite{PHENIX:2022lgn}. Figure~\ref{fig:STAR_DijetALL} shows that the STAR dijet data ~\cite{STAR:2021mqa} also strongly disfavors distributions with large and negative gluon helicities. In the plot the asymmetries $A_{LL}$ are presented for four dijet event topologies, namely, with forward-forward jets (top left), forward-central jets (top right), central-central jets (bottom left), and forward-backward jets (bottom right), where forward jet rapidity is $0.3<\eta<0.9$, central jet rapidity is $|\eta| < 0.3$, and backward jet rapidity is $-0.9 < \eta < -0.3$. The forward-forward and forward-central configurations probe the most asymmetric collisions down to $x \simeq 0.015$. The forward-forward and central-central events probe collisions with $|\cos\theta^*|$ near zero, whereas forward-central and forward-backward events are more sensitive to larger $|\cos\theta^*|$, where $\theta^*$ is the scattering angle in the center-of-mass frame of scattering partons. In both Figs.~\ref{fig:PHENIX_PhotonALL} and~\ref{fig:STAR_DijetALL}, the DSSV14 calculations are plotted as the black curves with the $1\sigma$ uncertainty bands marked in light blue. The blue curves with $1\sigma$ uncertainty bands in dark blue show the impact of all the data sets included in the new preliminary DSSV fit \cite{DSSV:prel} as in Fig.~\ref{fig:Gluon-impact}. The curves for JAM $\Delta g<0$ solution \cite{Zhou:2022wzm} are presented in red.

\subsection{Nonlinear QCD effects}
To understand where the saturation of gluon densities sets in, whether there is a simple boundary that separates this region from that of more dilute quark-gluon matter, is one of the most important physics cases of the RHIC Cold QCD program and future EIC. 

It is well known that PDFs grow rapidly at small-$x$. The power-law growth of the gluon density can be explained by gluon splitting, which leads to a linear evolution of gluon dynamics.
But if one imagines how such a high
number of small-$x$ partons would fit in the (almost) unchanged proton radius, one arrives at the
picture that the gluons and
quarks are packed very tightly in the transverse
plane. The typical distance between the partons
decreases as the number of partons increases, and
can get small at low-$x$ (or for a large nucleus instead of the proton). 
In QCD, the black disk limit states that the total hadronic cross section cannot grow forever. Thus, the growth of gluon density which can be described by BFKL evolution, has to be tamped at some point. At very high density, partons may start to recombine with each other on top of splitting. The recombination of two partons into one is proprtional to the number of pairs of the partons.
Therefore the BFKL function needs to be modified by adding a nonlinear term of recombination. Saturation is a new regime of QCD, where gluon splitting and recombination reach a balance.
One can define the saturation
scale as the inverse of this typical transverse interparton distance. Hence $Q_{s}$ indeed grows with $A$ and decreasing $x$.

Collisions between hadronic systems, $i.e.$, $pA$ and $dA$ at RHIC provide a window to the parton distributions of nuclei at small momentum fraction $x$ (down to $10^{-3}$). Several RHIC measurements have shown that,
at forward pseudorapidities (deuteron going direction),
the hadron yields are suppressed in $dAu$ collisions relative to $pp$ collisions in inclusive productions~\cite{BRAHMS:2003sns,Arsene:2004ux,Adler:2004eh,Adams:2006uz} and di-hadron correlations~\cite{Adams:2006uz,Adare:2011sc}. However, for the inclusive channel, it was indicated that the nuclear modified fragmentation can serve as another interpretation beyond gluon saturation to explain the suppression. 
The di-hadron correlation measurement can provide future test for the saturation physics. For di-hadron correlation in $dA$, the contributions from double-parton scatterings (DPS) to the $d$+$A$$\rightarrow \pi^0 \pi^0$X cross section are suggested as an alternative explanation for the suppression~\cite{Strikman:2010bg} beyond gluon saturation. Therefore, it is important to make the same measurements in the theoretically and experimentally cleaner $pA$ collisions.
Under the color glass condensate (CGC) framework~\cite{McLerran:1993ni,McLerran:1993ka,iancu2002colour}, 
at a given $x$, gluons from different nucleons are predicted to amplify the total transverse gluon density by a factor of $A^{1/3}$ for a nucleus with mass number $A$. RHIC 2015 $pp$, $pAl$, and $pAu$ datasets are ideal to study the $A$-dependence by varying the nuclei species.

\begin{figure*}[tbh]
\centering
\includegraphics[width=0.45\textwidth]{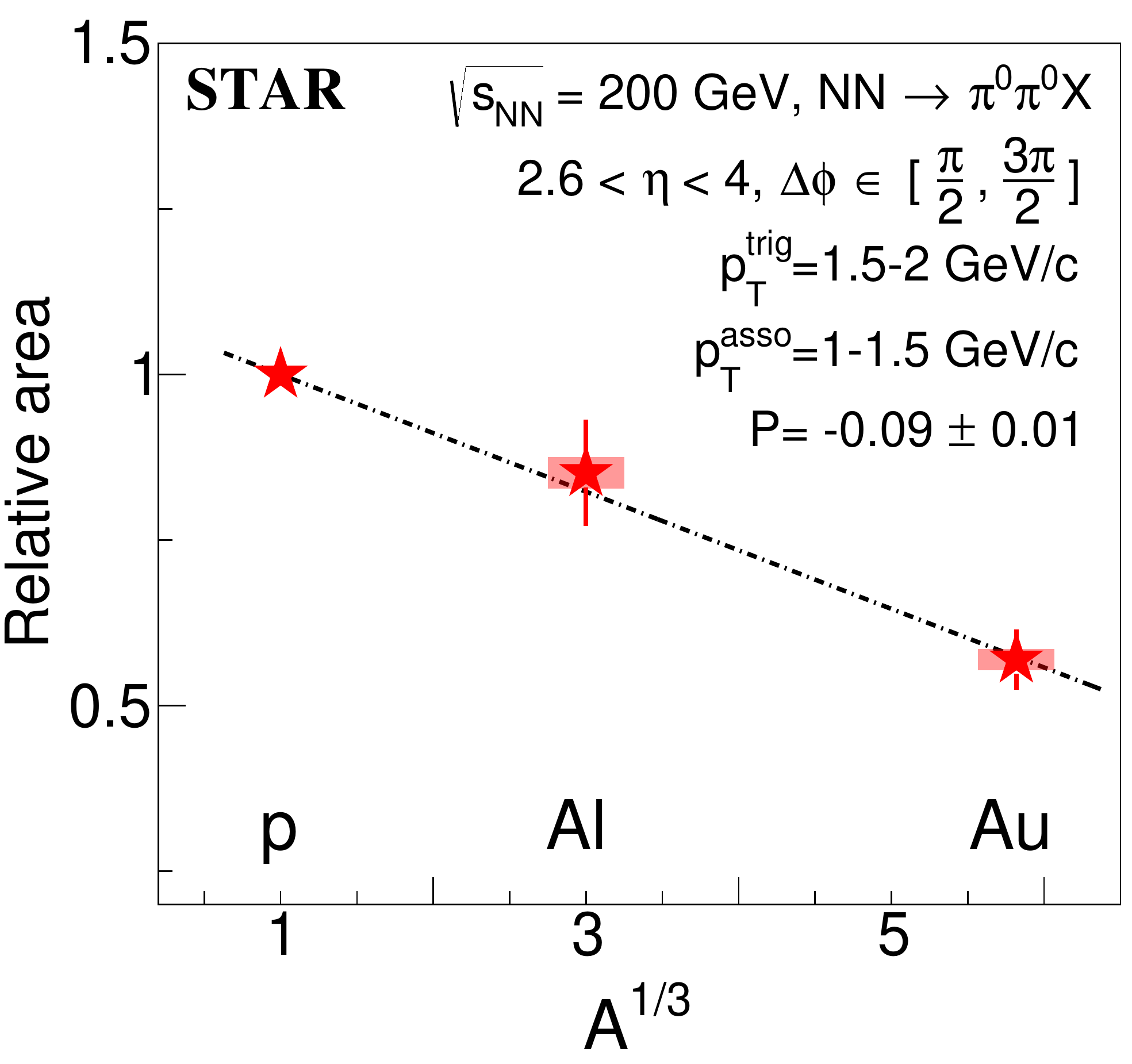}
\includegraphics[width=0.45\textwidth]{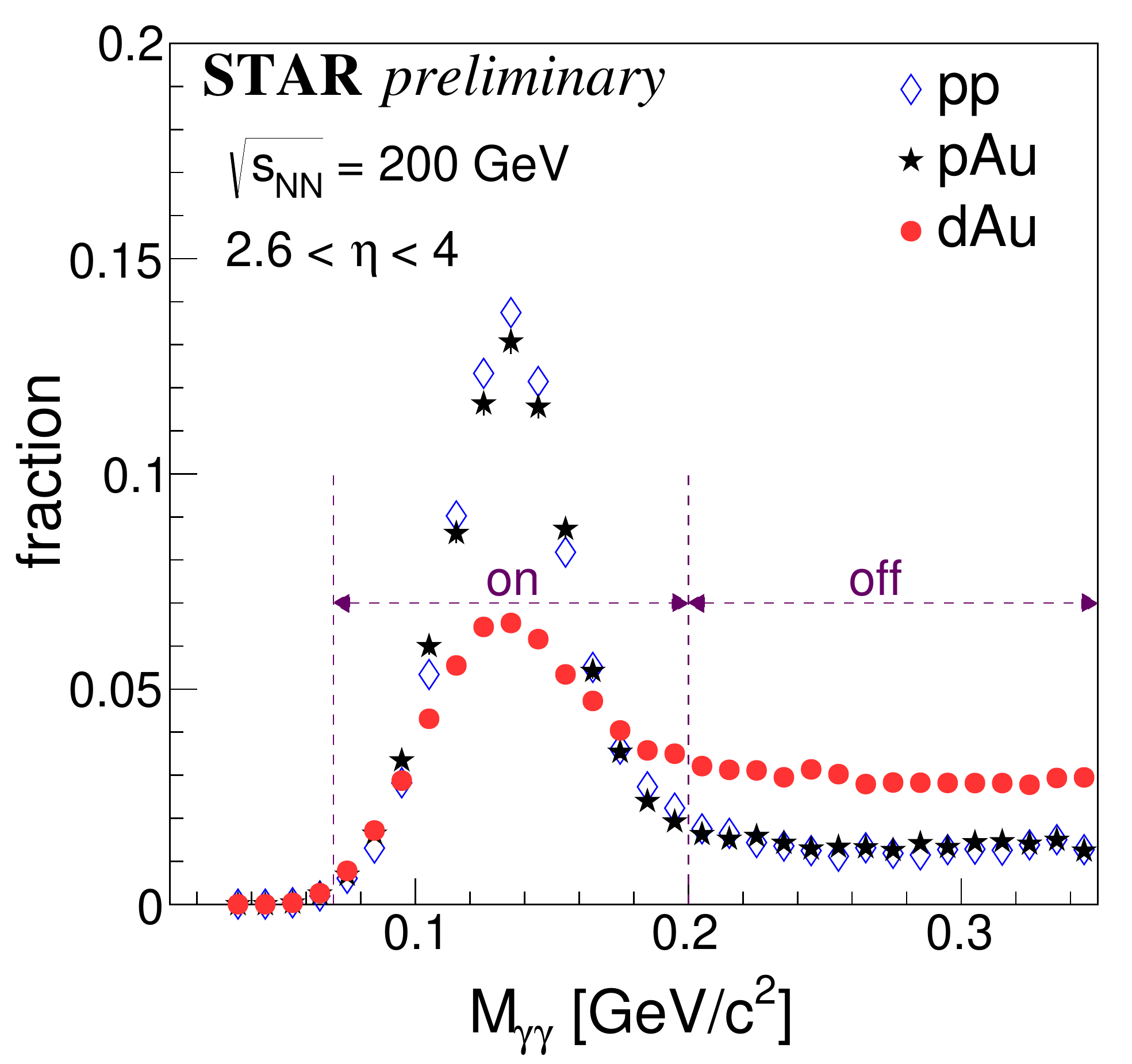}
\caption{ Left: At $p_{T}^{trig}$ = 1.5$-$2 GeV/$c$ and $p_{T}^{asso}$ = 1.5$-$2 GeV/$c$, a linear dependence of the suppression of back-to-back $\pi^{0}$ correlation as a function of $A^{1/3}$ is observed within the uncertainties, the slope ($P$) is found to be $-0.09$ $\pm$ $0.01$. The plot is from \cite{STAR:2021fgw}.
Right: The reconstructed invariant mass of two photons for $pp$, $pAu$, and $dAu$ collisions. The background is higher in $dAu$ collisions in comparison with the $pp$ and $pAu$ collisions. $pp$ and $pAu$ collisions are similar.}
\label{fig:dipi0pppAdA}
\end{figure*}

\begin{comment}

\begin{SCfigure}[][h]
    \centering
    \includegraphics[width=0.5\columnwidth]{fig/dipi0_A}
    \caption{At $p_{T}^{trig}$ = 1.5$-$2 GeV/$c$ and $p_{T}^{asso}$ = 1.5$-$2 GeV/$c$, a linear dependence of the suppression of back-to-back $\pi^{0}$ correlation as a function of $A^{1/3}$ is observed within the uncertainties, the slope ($P$) is found to be $-0.09$ $\pm$ $0.01$. The plot is from \cite{STAR:2021fgw}.}
    \label{fig:dipi0pA}
\end{SCfigure}

\begin{SCfigure}[][h]
    \centering
 \includegraphics[width=0.5\columnwidth]{fig/pi0mass_pppaudau}
\caption{The reconstructed invariant mass of two photons for $pp$, $pAu$, and $dAu$ collisions. The background is higher in $dAu$ collisions in comparison with the $pp$ and $pAu$ collisions. $pp$ and $pAu$ collisions are similar.
}\label{fig:dipi0dAu}
\end{SCfigure}
\end{comment}

The recent published forward di-$\pi^{0}$ correlation measured by the STAR detector pioneered the observation of the dependence of nonlinear gluon dynamics on the nuclear mass number $A$ \cite{STAR:2021fgw}, see the left panel of Fig. \ref{fig:dipi0pppAdA}. 
The area is extracted by a Gaussian fit of the back-to-back correlation measured from each collision system. The area ratio of $pA/pp$ presents the relative yields of back-to-back di-$\pi^{0}$s in $pA$ with respect to $pp$ collisions.
The area ratio in $pAu$ over $pp$ is about 50\% indicating a clear suppression of back-to-back di-$\pi^{0}$ correlation in $pAu$ compared to $pp$ collisions. The same trend but smaller amount of suppression is observed in $pAl$ collisions. The suppression is found to scale with $A$ and linearly dependent on $A^{1/3}$.
The extracted slope from the linear dependence will be critical input for the gluon saturation model in CGC.
Meanwhile, STAR revisited the same measurement for $dAu$ collisions. It was predicted by comparing the forward di-$\pi^{0}$ correlation in $pp$, $pAu$, and $dAu$ collisions, one can access the contribution from DPS \cite{Strikman:2010bg}.

%STAR revisited the same di-$\pi^{0}$ correlation measurement in $dAu$ collisions, aiming at resolving the question whether there is effect from DPS.
For RHIC 2016 data, a large background of $\pi^{0}$ identification is found in $dAu$ collisions, in comparison with the $pp$ and $pAu$ collisions from 2015 in the right panel of Fig. \ref{fig:dipi0pppAdA}. The generated combinatoric correlation dominates in $dAu$ collisions, which makes it very challenging to identify the signal correlation. The forward di-$\pi^0$ correlation measurement favors the cleaner $pA$ collisions rather than $dA$ collisions. It emphasizes the importance of measuring the di-hadron correlation in $pA$ collisions with the STAR Forward Upgrade in the future run 2024. The higher delivered integrated luminosity for this run together with the Forward Upgrade will enable one to study more luminosity-hungry processes and/or complementary probes to the di-$\pi^0$ correlations, i.e. di-hadron correlations for charged hadrons, photon-jet, photon-hadron and di-jet correlations. Utilizing the forward tracking systems, the background for particle identification will be much suppressed with respect to the current di-$\pi^0$ studies.

These results are crucial for the equivalent measurements at the EIC, which are planned at close to identical kinematics, because only if non-linear effects are seen with different complementary probes, i.e., $ep$ and $pA$, one can claim a discovery of saturation effects and their universality.  Therefore it is imperative that analysis activities related to the unpolarized Cold-QCD program continue to be supported throughout the upcoming years. 

%\end{multicols}

%% file: 3DImage.tex
\label{3DImageChapter}
\begin{SCfigure}
    \centering
    \includegraphics[width=0.5\columnwidth]{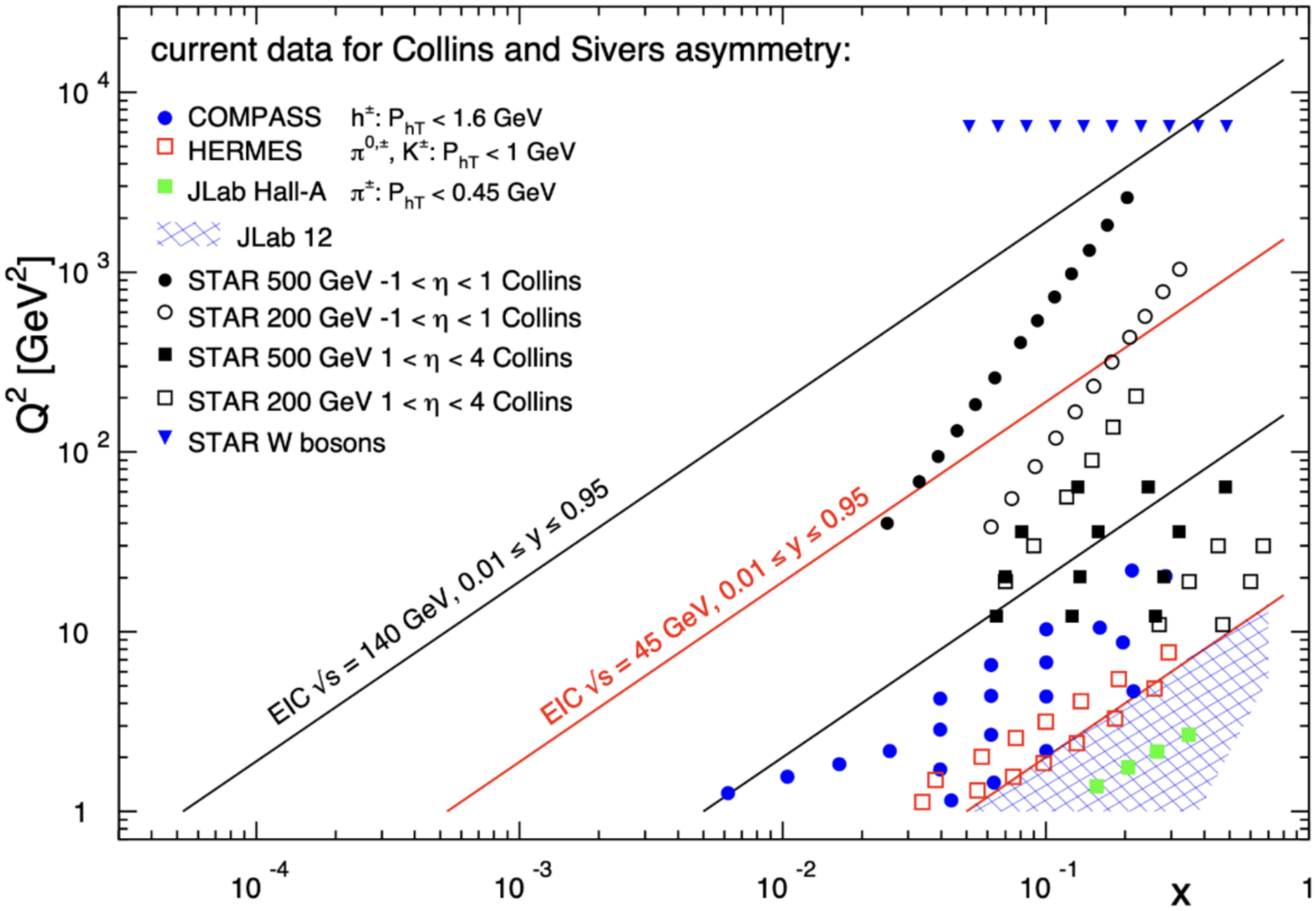}
    \caption{ The $x$-$Q^{2}$ probed with data from the future
EIC and Jlab-12 GeV as well as the current SIDIS
data and the jet and $W$-boson data from RHIC. All data are
sensitive to the Sivers function and transversity
times the Collins FF in the TMD formalism.}
    \label{fig:spin-xQ2transverse}
\end{SCfigure}

STAR and PHENIX opened new territory in studying the 3D structure of the proton in the region of momentum fractions down to $x \sim$ 0.01 and high $Q^{2}$, a region not probed by prior experiments. See Fig.~\ref{fig:spin-xQ2transverse}. 

%The impact of STAR 2011 W/Z data can be found in Fig. \ref{fig:WZ_AN_impact_PV}, \ref{fig:WZ_AN_impact_Up}, and \ref{fig:WZ_AN_impact_Down}. The impact of STAR forward EM jet data can be found in Fig.~\ref{fig:EMjet_AN}.
%Fig.~\ref{fig:EMjet_impact}.

\begin{itemize}
\item The collected unique sets of transversely polarized data in $pp$ and $pA$ collisions, including the most recent campaign with the forward upgrade, will be finalized with the 2024 RHIC run by STAR and sPHENIX. 

\item To accomplish the scientific mission of the transverse spin program, it is imperative that analysis activities continue to be supported throughout the upcoming years. These activities offer discovery potential of their own, and they are critical for properly interpreting data from the future Electron-Ion Collider.

\item STAR pioneered the novel use of jets and their substructure to study initial and final state transverse momentum dependent (TMD) effects in polarized {\pp} collisions. For example, the measured single-spin asymmetries of hadrons in jets probe the quark transversity distribution and Collins TMD fragmentation function, and the single-spin asymmetry of dijet opening angle is sensitive to the Sivers TMD parton distribution.

\item  STAR has also measured quark transversities via dihadron interference fragmentation functions.  The results from early measurements have been included in a global analysis, and found to provide significant constraints.  Ongoing analysis of more recent STAR data, together with the data that STAR will record during 2024, will provide far more stringent constraints.

\item Substantial progress on the large forward transverse single-spin asymmetry puzzle has been made. The $A_N$ of the isolated $\pi^0$s was found to be significantly larger than that for non-isolated ones both in {\pp} and $pA$ collisions at STAR.  The $A_N$ for $\pi^0$s at large $x_F$, far forward pseudorapidity ($\eta>6)$, and $p_T<1$ GeV/$c$ at RHICf was found to be comparable to that at the same $x_F$, but with $2.5<\eta<4$ and $p_T>2$ GeV/$c$ at STAR.  The $A_N$ for electromagnetic jets was found to be small but non-zero, which provided significant constraints to the quark Sivers function. The $A_N$ for forward diffractive EM-jets has been measured and found not to be the source of the large $A_N$.  In fact, it favors a negative contribution.

\item Transverse single-spin asymmetry $A_{N}$ of weak bosons, sensitive to the Sivers TMD function, has been probed at STAR. With the increased precision provided by 2017 data, STAR found smaller asymmetries than were suggested by 2011 data. As a result, the increased statistics of the 2022 dataset are critical to improve the precision of our asymmetry measurements in order to provide a conclusive test of the Sivers' function sign change.
%\item W/Z $A_N$ and unpolarized TMD measure (Xiaoxuan)

\item PHENIX has measured transverse single-spin asymmetries at mid-rapidity that provide constraints on the twist-3 correlation functions of quarks and gluons, including the first RHIC result of direct photon $A_N$, open heavy flavor decay electron $A_N$, and high precision neutral meson $A_N$.

\item PHENIX and STAR have both measured the nuclear dependence of the forward inclusive hadron single-spin asymmetries.  PHENIX finds a strong nuclear dependence for positive hadrons at $1.2<\eta<2.4$, whereas STAR finds a weak nuclear dependence for $\pi^0$ at $2.7<\eta<3.8$.  Neither the origin of the nuclear dependence, nor the difference between the PHENIX and STAR results is well understood at this time.

\item Transverse single-spin asymmetry of exclusive $J/\psi$ photoproduction in ultra-peripheral collisions is expected to directly probe the generalized parton density (GPD) distribution. The STAR forward detector and data beyond 2022 can measure unique kinematic phase space, e.g., close to the threshold production energy of $J/\psi$, where a large asymmetry signal is expected.

\end{itemize}

\newpage
%\begin{multicols}{2}
\subsection{Studies of initial and final state TMD effects with jets}

%JLD: Do we need this first sentence (looks like a copy-and-paste of bullet point from earlier), or should we just dive right into the paragraph on Collins?
%CAG:  I agree that it's a copy-paste.  However, I still think it is useful to set the stage.
STAR has pioneered the novel use of jets and their substructure to study initial state and final state TMD effects in polarized {\pp} collisions.

\begin{SCfigure}[][h]
    \centering
    \includegraphics[width=0.5\columnwidth]{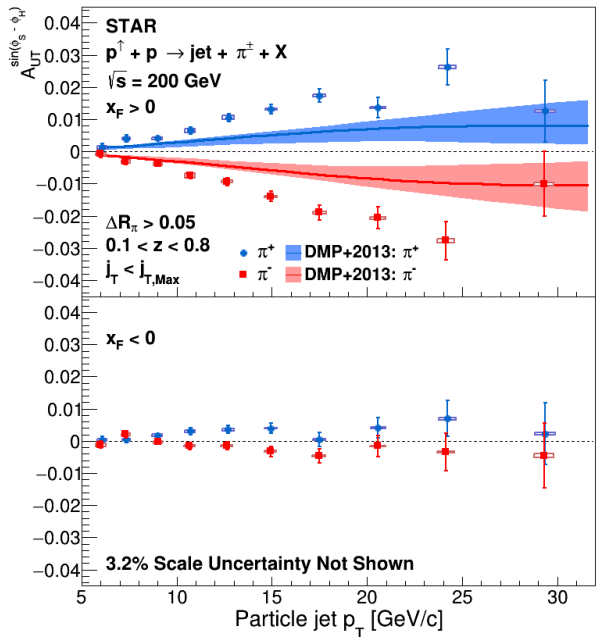}
    \caption{Collins asymmetry plotted for identified $\pi^{+}$ (blue) and $\pi^{-}$ (red) particles as a function of jet $p_{T}$ for jets that scatter forward relative to the polarized beam ($x_{F}>0$) in the top panel and those that scatter backward ($x_{F}<0$) in the lower panel, extracted from data collected in 2012 and 2015 \cite{STAR:2022hqg}. The full ranges of both $z$ and $j_{T}$ are integrated over. Theoretical evaluations from \cite{DAlesio:2017bvu} with their uncertainties are presented for $\pi^{+}$ (blue) and $\pi^{-}$ (red).
    Source: \cite{STAR:2022hqg}.
    }
    \label{fig:figure_CollinsAsym_vs_jetPt}
\end{SCfigure}

\begin{SCfigure}[][h]
     \centering
     \includegraphics[width=0.5\columnwidth]{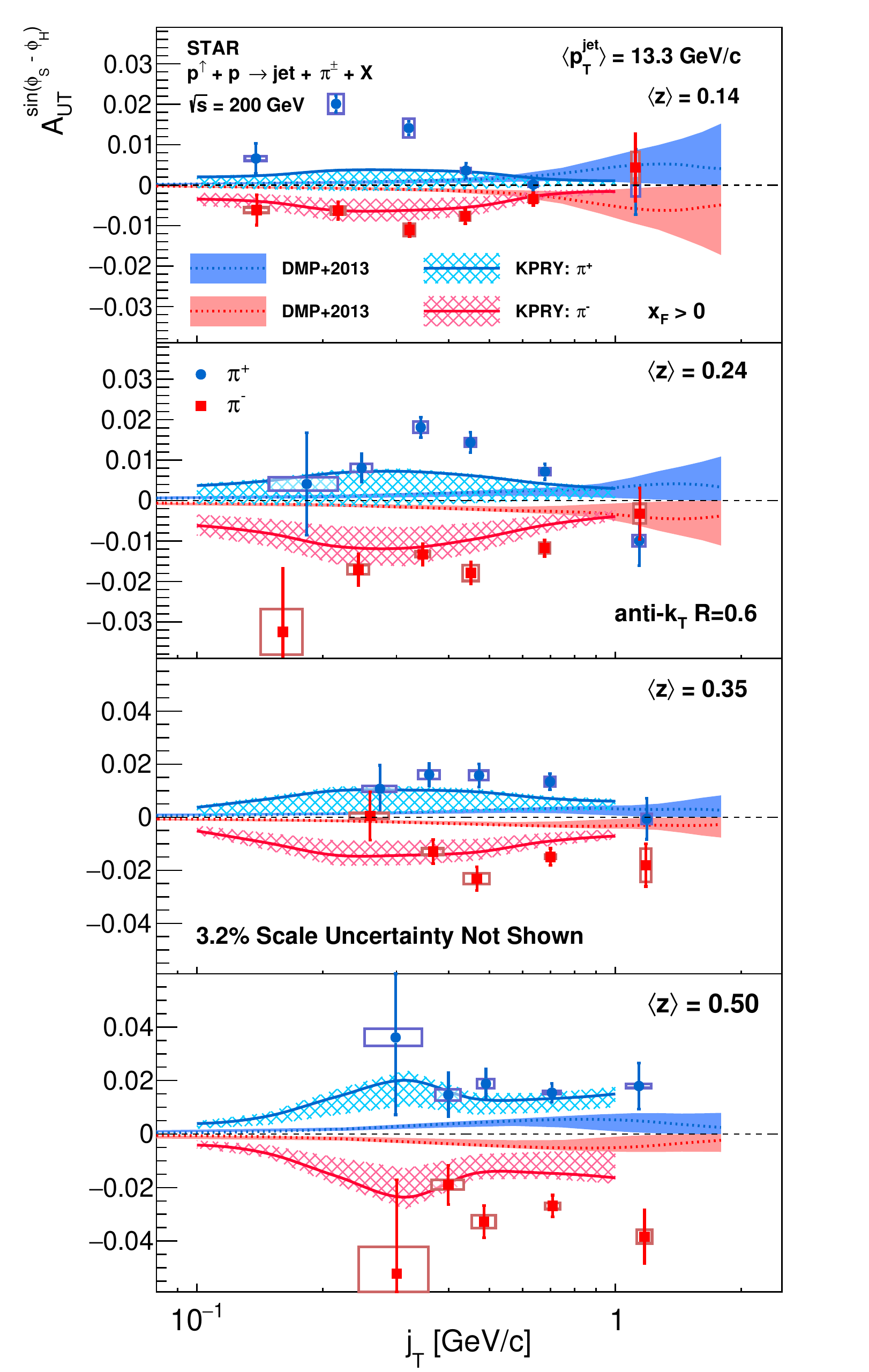}
     \caption{Collins asymmetry plotted for identified $\pi^{+}$ (blue) and $\pi^{-}$ (red) particles as a function of $j_T$ for four separate bins of hadron $z$, in jets with $p_T > 9.9$ GeV/$c$ and $0 < \eta < 0.9$.  Theoretical evaluations from \cite{Kang:2017btw} and \cite{DAlesio:2017bvu} are also shown.
     Source:  \cite{STAR:2022hqg}.
     }
     \label{fig:Collins_figure_z_jT}
\end{SCfigure}

The single-spin asymmetries of the azimuthal distribution of identified pions, kaons, and protons in high-energy jets measured at STAR probe the \textit{collinear} quark transversity in the proton, coupled to the transverse momentum dependent Collins fragmentation function \cite{Kang:2017glf,Kang:2017btw,DAlesio:2010sag}. This makes \pp\ collisions a more direct probe of the Collins fragmentation function than SIDIS, where a convolution with the TMD transversity distribution enters. The Collins asymmetry in \pp\ collisions is an ideal tool to explore the fundamental QCD questions of TMD factorization, universality, and evolution. Figure~\ref{fig:figure_CollinsAsym_vs_jetPt} shows the recent results on combined 2012 and 2015 Collins asymmetries for charged pions within jets as a function of jet $p_{T}$ \cite{STAR:2022hqg}.  By integrating over the  hadron longitudinal and transverse momenta within the jets, Fig.~\ref{fig:figure_CollinsAsym_vs_jetPt} is sensitive primarily to the quark transversity. The measured asymmetries for jets that scatter forward relative to the polarized beam are larger than theoretical predictions~\cite{DAlesio:2017bvu}, which are based on the transversity and Collins fragmentation function from SIDIS and $e^{+}e^{-}$ processes within the TMD approach. Alternatively, the asymmetries can be investigated as functions of $z$, the fraction of jet momentum carried by the hadron, and $j_{T}$, the momentum of the pion transverse to the jet axis, as shown in Fig.~\ref{fig:Collins_figure_z_jT}.  This provides a direct measurement of the kinematic dependence of the Collins fragmentation function. The $j_T$ dependence appears to vary with $z$, contrary to the assumptions of most current phenomenological models~\cite{Kang:2017glf,Kang:2017btw,DAlesio:2010sag}. STAR has also published Collins asymmetry measurements from a smaller 500 GeV data set collected in 2011~\cite{STAR:2017akg}. While statistics are limited, the results are consistent with those at 200 GeV for overlapping $x_T$, despite sampling $Q^2$ that is larger by a factor of 6. Analysis of the higher statistics 510 GeV data collected in 2017 is underway and will provide unique insight into the $Q^2$ evolution of the Collins TMD fragmentation function.
Concurrent with the Collins effect measurements, STAR has also measured azimuthal modulations that are sensitive to the twist-3 analogs of the quark and gluon Sivers functions and to linear polarization of gluons in transversely polarized protons~\cite{STAR:2017akg,STAR:2022hqg}.  Analysis is also underway to determine the unpolarized TMD fragmentation functions.

\begin{SCfigure}[][h]
  \centering
  \includegraphics[width=0.5\columnwidth]{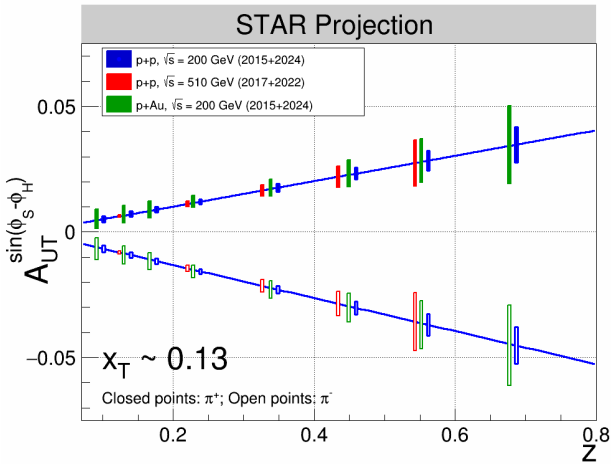}
  \caption{Projected statistical uncertainties for STAR Collins asymmetry measurements at $0 < \eta < 0.9$ in \pp\ at $\sqrt{s}$ = 200 and 510 GeV and $pAu$ at \snn\ = 200 GeV. The points have arbitrarily been drawn on the solid lines, which represent simple linear fits to the STAR preliminary 200 GeV \pp\ Collins asymmetry measurements from 2015. (Note that only one bin is shown spanning $0.1 < z < 0.2$ for 510 GeV $pp$, whereas three bins are shown covering the same $z$ range for the 200 GeV measurements.)}
  \label{fig:Collins_proj_mid}
\end{SCfigure}

As shown in Fig.~\ref{fig:spin-xQ2transverse}, data from 200 GeV \pp\ collisions from the upcoming run 2024 with the STAR Forward Upgrade will interpolate between the coverage that we will achieve with the forward data collected at 510 GeV in 2022 at high-$x$ and the data at low-$x$ from the STAR mid-rapidity detectors. Overall, all STAR data will provide valuable information about evolution effects and, with the projected statistical precision presented in Fig.~\ref{fig:Collins_proj_mid}, will establish the most precise benchmark for future comparisons to $ep$ data from the EIC. It is also important to recognize that the hadron-in-jet measurements with the STAR Forward Upgrade will provide a very valuable experience detecting jets close to beam rapidity that will inform the planning for future jet measurements in similar kinematics at the EIC. 

STAR also has the unique opportunity to extend the Collins effect measurements to nuclei. This will provide an alternative look at the universality of the Collins effect in hadron production (by dramatically increasing the color flow options of the sort that have been predicted to break factorization for TMD PDFs like the Sivers effect~\cite{Collins:2007,Rogers:2010}) and explore the spin dependence of the hadronization process in cold nuclear matter. STAR collected a proof-of-principle dataset during the 2015 $pAu$ run that is currently under analysis. Those data will provide the first estimate of medium-induced effects. However, the small nuclear effects seen by STAR for forward inclusive $\pi^0$ $A_N$~\cite{STAR:2020grs} indicate that greater precision will likely be needed. Figure~\ref{fig:Collins_proj_mid} shows the projected statistical uncertainties for the $pAu$ Collins asymmetry measurement at \snn\ = 200 GeV from 2015 and 2024 data, compared to those for \pp\ at the same energy.

\begin{SCfigure}[][h]
  \centering
\includegraphics[width=0.5\textwidth]{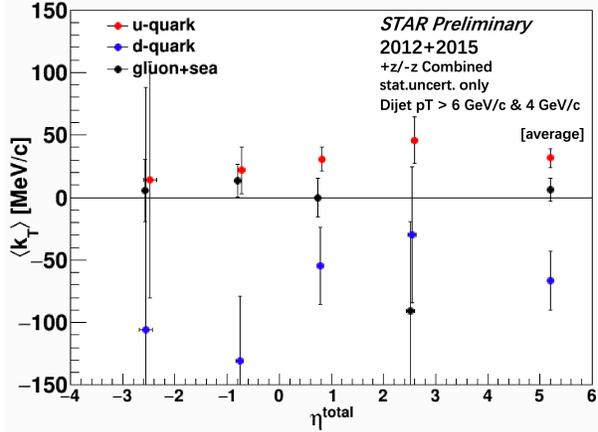}
  \caption{Preliminary results of the average transverse momentum $\left<k_T\right>$ for individual partons, inverted using parton fractions from simulation and tagged $\left<k_T\right>$ in data, plotted as a function of summed pseudorapidities of the outgoing jets $\eta_{\mathrm{total}} \sim \log(x_1/x_2)$.  (Positive $\eta_{\mathrm{total}}$ represents dijets emitted in the direction of the polarized beam.) The rightmost points represent the average of all the $\eta_{\mathrm{total}}$ bins. The systematic
uncertainty in $\eta_{\mathrm{total}}$ is set to be non-zero to improve the visibility of the error bars. Source: \cite{stardijetSiversAsym20122015}.}
  \label{fig:Sivers-Dijet-200-PRL-Draft}
\end{SCfigure}

Another example of utilizing jets to unravel the internal TMD structure of the proton is the measurement of the asymmetry of the spin-dependent `tilt' of the dijet opening angle, which is sensitive to the Sivers TMD PDF. For transversely polarized protons, the Sivers effect probes whether the transverse momentum $\vec{k}_T$ of the constituent quarks is preferentially oriented in a direction perpendicular to both the proton momentum and its spin.  Figure~\ref{fig:Sivers-Dijet-200-PRL-Draft} shows the first-ever observation of the Sivers effect in dijet production from the 200 GeV transverse spin data that STAR recorded in 2012 and 2015 \cite{stardijetSiversAsym20122015}. The jets are sorted according to their net charge $Q$, yielding jet samples with enhanced contributions from $u$ quarks (positive $Q$) and $d$ quarks (negative $Q$), with a large set near $Q=0$ dominated by gluons. Simple kinematics allow for conversion from the spin-dependent `tilt' of the dijet pair to a value of $k_T$ on an event-by-event basis. Finally, the results are unfolded for the $k_T$ of individual partons. Such measurements are crucial to explore questions regarding factorization of the Sivers function in dijet hadroproduction \cite{Collins:2007,Rogers:2010,Liu:2020jjv,Kang:2020xez}. New data to be taken in 2024 will reduce the uncertainties for the region of summed pseudorapidities of the outgoing jets $|\eta_3+\eta_4|<1$ by about a factor of two. The increased acceptance from the iTPC will reduce the uncertainties at $|\eta_3+\eta_4| \approx 2.5$ by a much larger factor, while the Forward Upgrade will enable the measurements to be extended to even larger values of $|\eta_3+\eta_4|$. When combined with the 510 GeV data from Run-17 and Run-22, the results will provide a detailed mapping \textit{vs}.\@ $x$ for comparison to results for Sivers functions extracted from SIDIS, Drell-Yan, and vector boson production.

\subsection{Transversity from di-hadron interference fragmentation functions}

\begin{SCfigure}[][h]
    \centering
    \includegraphics[width=0.5\textwidth]{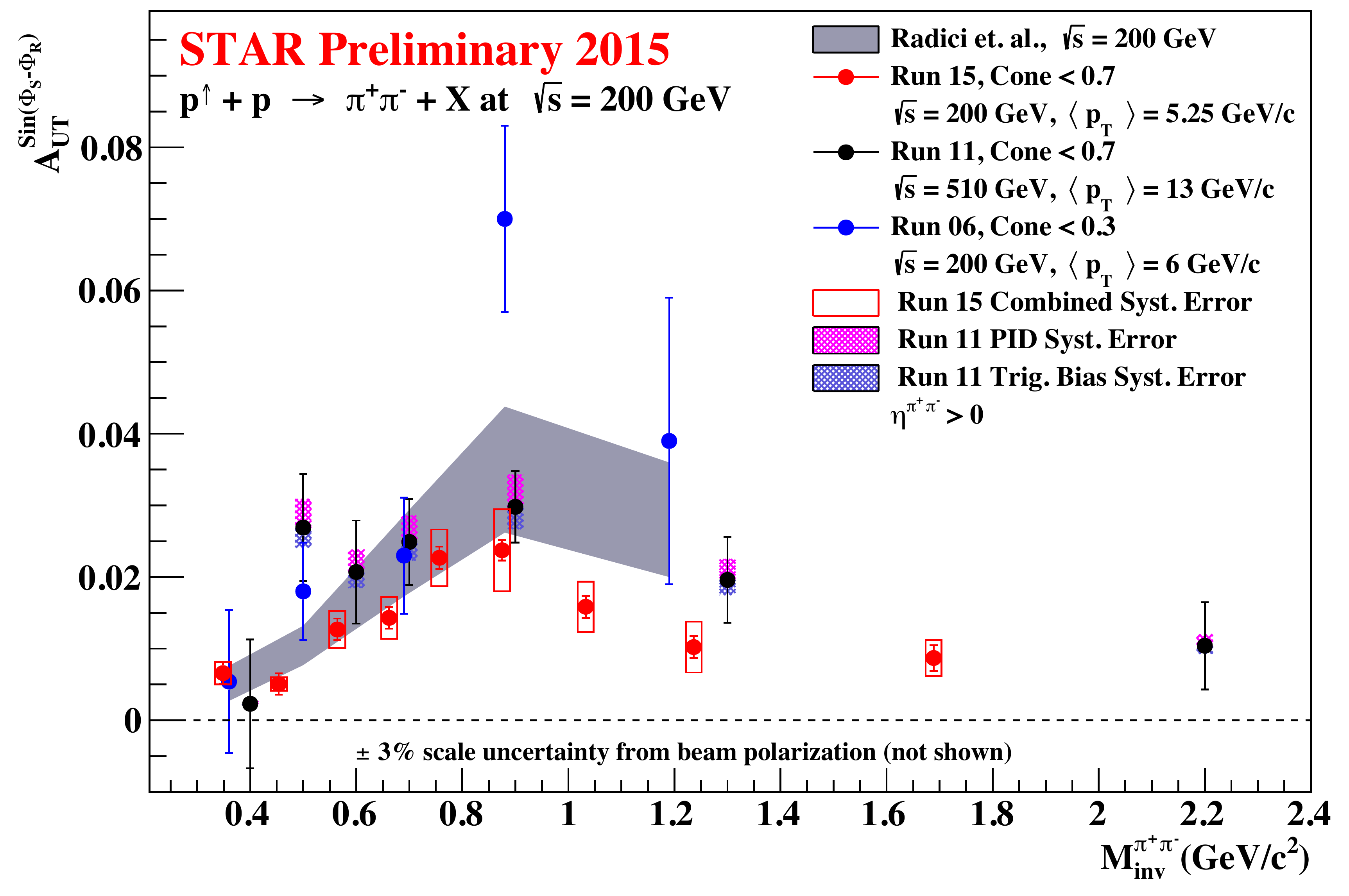}
    \caption{A comparison of STAR published \cite{STAR:2015.115.242501,STAR:2018.780.332} and preliminary \cite{starIFF} IFF asymmetries vs. dipion invariant mass to predictions from the global analysis of \cite{Radici:2018.120.192001}, which only included the 200 GeV data from 2006 in the fit.  The $p_T$ bins at 200 and 500 GeV have been chosen to sample similar values of $x_T = 2 p_T/\sqrt{s}$.
    Source: \cite{starIFF}.}
    \label{fig:STAR_IFF_results}
\end{SCfigure}

STAR has also measured quark transversity via dihadron Interference Fragmentation Functions (IFF) in 200 and 500 GeV \pp\ collisions \cite{STAR:2015.115.242501,STAR:2018.780.332}, as shown in Fig.~\ref{fig:STAR_IFF_results}.
The IFF is a collinear observable, so these measurements provide a complementary probe of transversity relative to the Collins asymmetry measurements that obeys different evolution equations.  The results from the first measurements at 200 GeV, which were based on data recorded during 2006 \cite{STAR:2015.115.242501}, have been included together with IFF measurements from SIDIS in a global analysis \cite{Radici:2018.120.192001} that is also shown in Fig.~\ref{fig:STAR_IFF_results}.  The STAR IFF measurements were found to provide significant additional constraints on the $u$- and $d$-quark transversities.  The dominant systematic uncertainties in the global analysis arose from the current lack of knowledge regarding the unpolarized gluon dihadron fragmentation functions.  Analysis of the unpolarized IFF function is underway at STAR that will help to reduce these uncertainties. The analysis of IFF asymmetries with more recent STAR data taken in 2017 and 2022 at 510 GeV, together with the data that STAR will record during 2024 at 200 GeV, will provide far more stringent constraints on quark transversities than have been obtained to date when they are included in future global analyses.

\subsection{Transverse single-spin asymmetry in the forward region}

\begin{figure*}
    \centering
\includegraphics[width=5cm,clip]{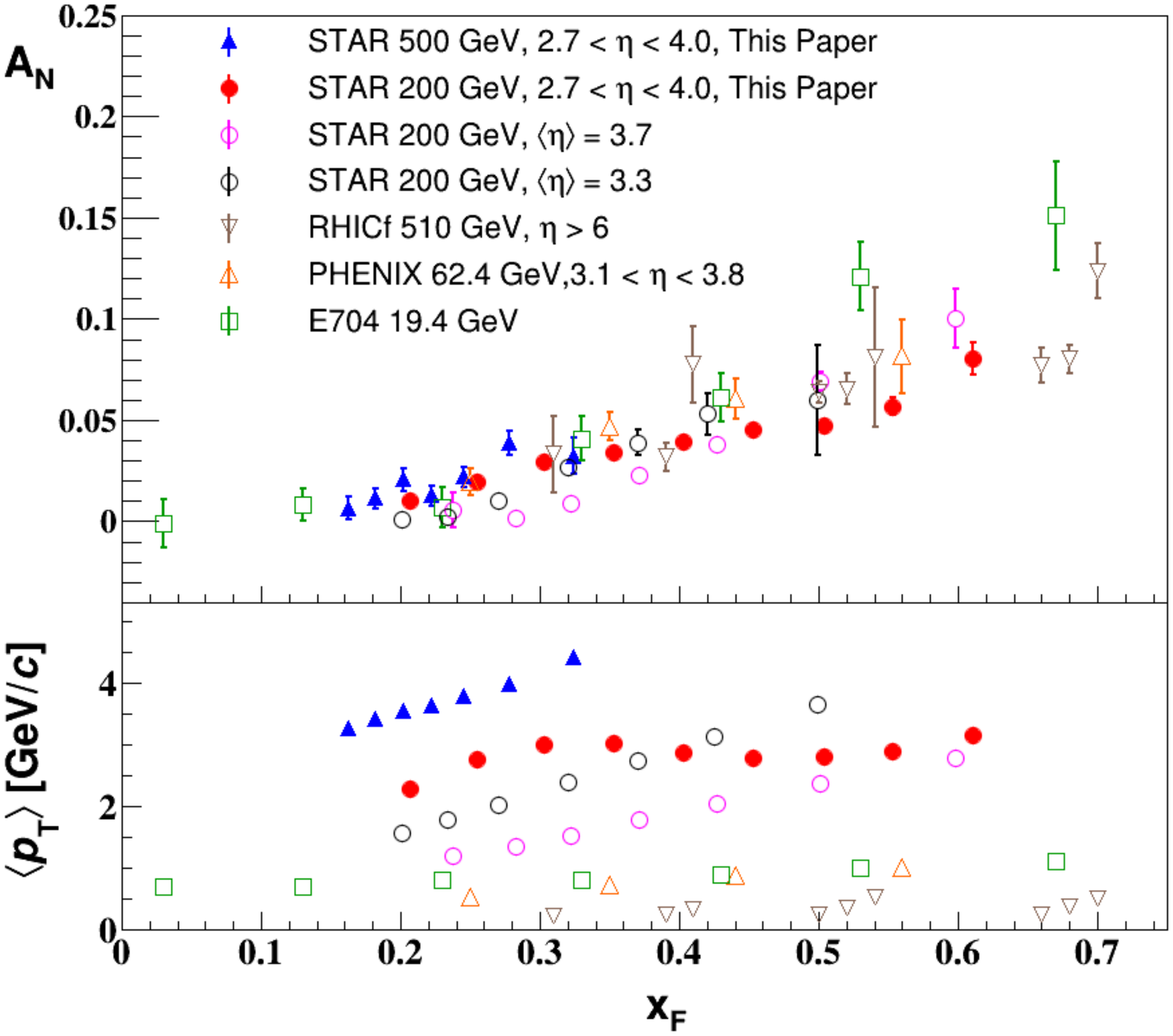} 
\includegraphics[width=5cm,clip]{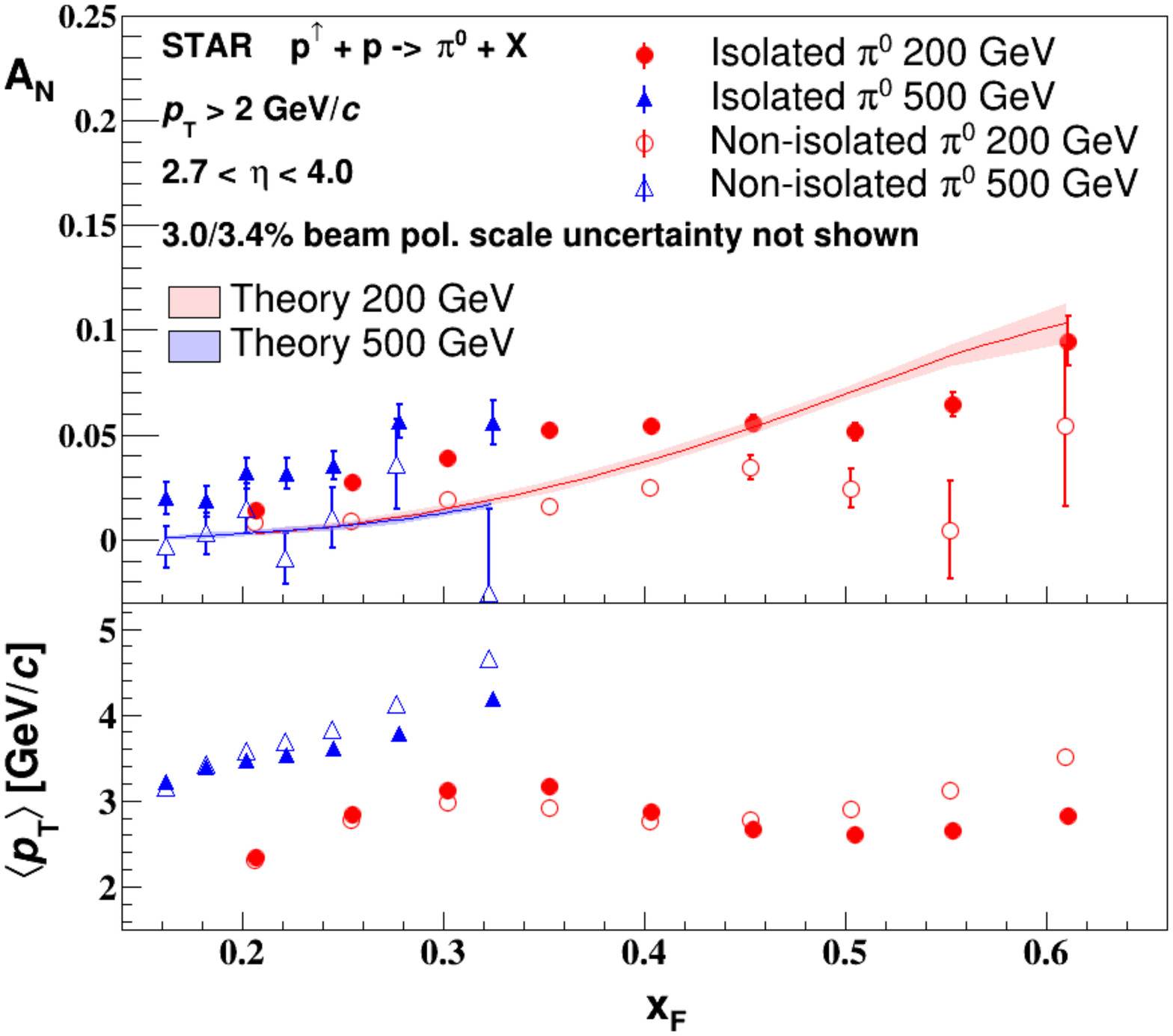}
\includegraphics[width=5.5cm,clip]{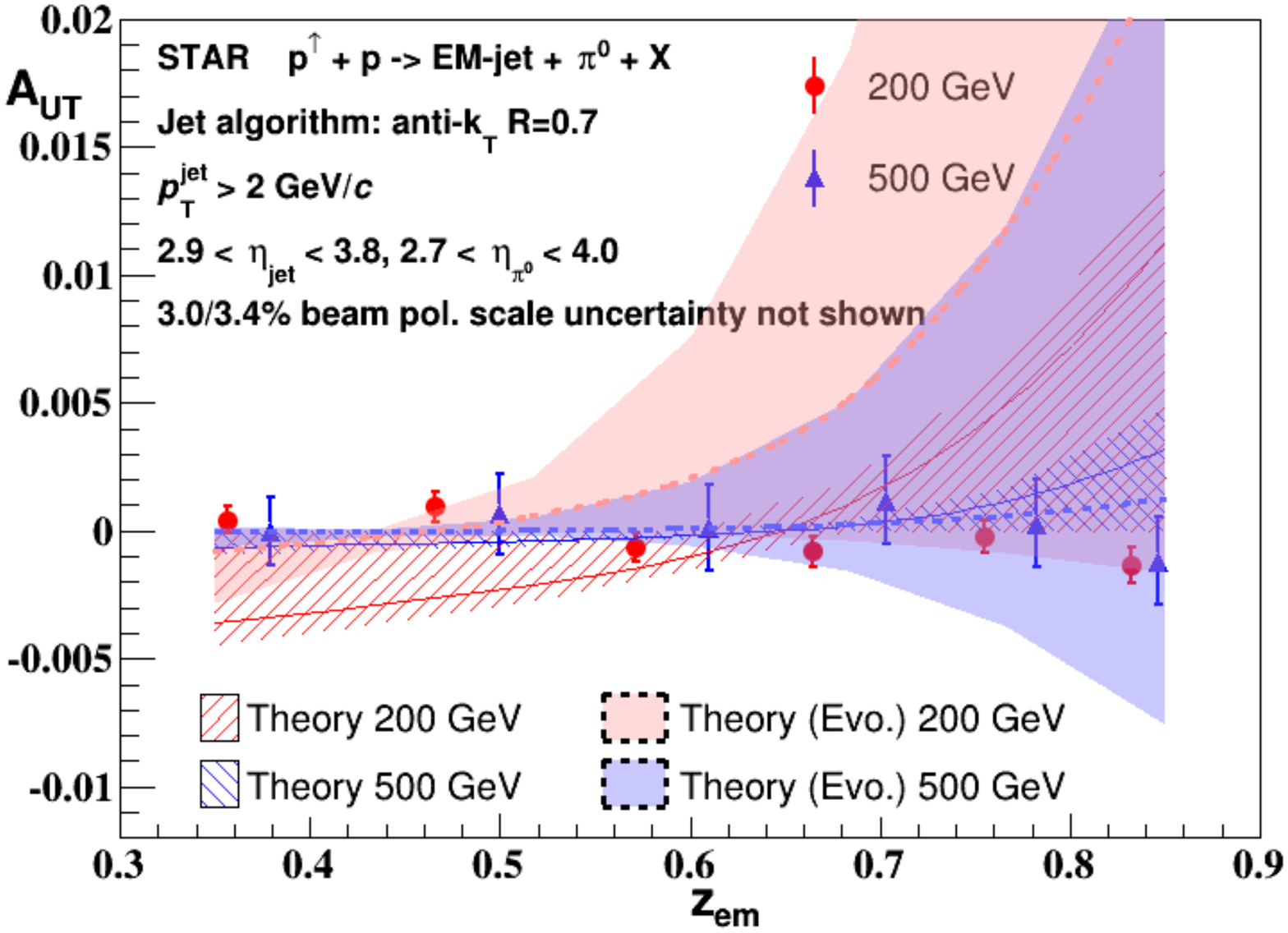} 
\caption{
Left: Transverse single-spin asymmetry $A_N$ as a function of $x_\mathrm{F}$ for inclusive $\pi^{0}$ in $pp$ collisions up to RHIC energies of 200 and 510 GeV. Middle: $A_N$ asymmetries for the isolated and non-isolated $\pi^{0}$ in $pp$ collisions at 200 and 500 GeV. Right: The Collins asymmetry for $\pi^{0}$ in an electromagnetic jet for $pp$ collisions at $\sqrt s=$ 200 and 500~GeV. The plots are from \cite{STAR:2020nnl}.
}
    \label{fig:pi0_AN}
\end{figure*}

STAR measurements have demonstrated the persistence of sizeable transverse single-spin asymmetries $A_N$ for forward $\pi^0$ production at RHIC energies up to 510 GeV with a weak energy dependence (see left panel of Fig. \ref{fig:pi0_AN}), where different QCD mechanisms including the high twist effect, TMD effects like the Sivers or Collins effects, and diffractive processes could all contribute. 
It is thus important to study different effects separately for a full understanding of the underlying mechanism, and a series of measurements were performed in $pp$ collsions at both 200 and 500 GeV and in $pA$ collisions at STAR~\cite{STAR:2020nnl,STAR:2020grs,STAR:DiffractiveDIS2022}. 

\begin{SCfigure}[][h]
    \centering
    \includegraphics[width=0.5\textwidth]{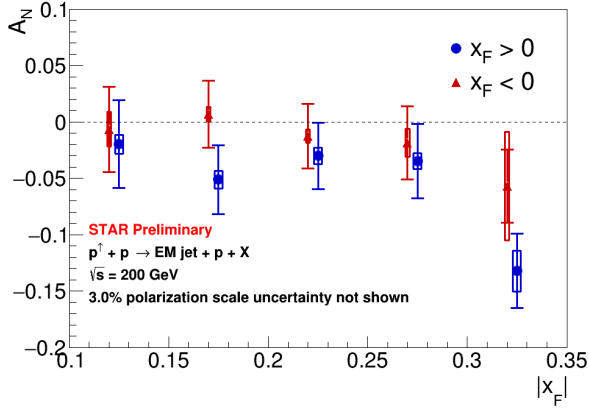}
    \caption{Transverse single-spin asymmetry for diffractive EM-jet as a function of $x_{F}$ in transversely polarized proton-proton collisions at $\sqrt{s}=$ 200 GeV~\cite{STAR:DiffractiveDIS2022}. The blue points are for $x_{F} > 0$. The red points are for $x_{F} < 0$ with a constant shift of -0.005 along x-axis for clarity. The rightmost points are for $0.3 < |x_{F}| < 0.45$. 
    %The systematic uncertainty (the rectangular boxes) for $A_{N}$ mainly comes from the event selections for suppressing the background events~\cite{STAR:DiffractiveDIS2022}.
    }
    \label{fig:diffractive EM-jet AN}
\end{SCfigure}

\begin{figure*}
    \centering
\includegraphics[width=7.2cm,clip]{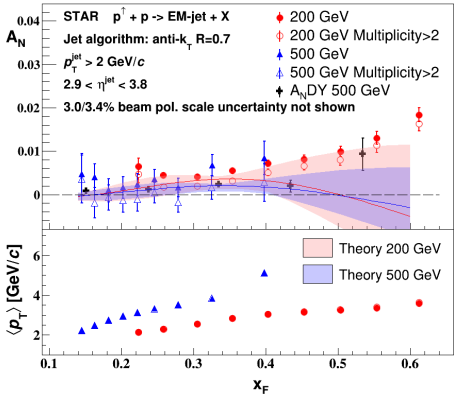} 
\includegraphics[width=7.9cm,clip]{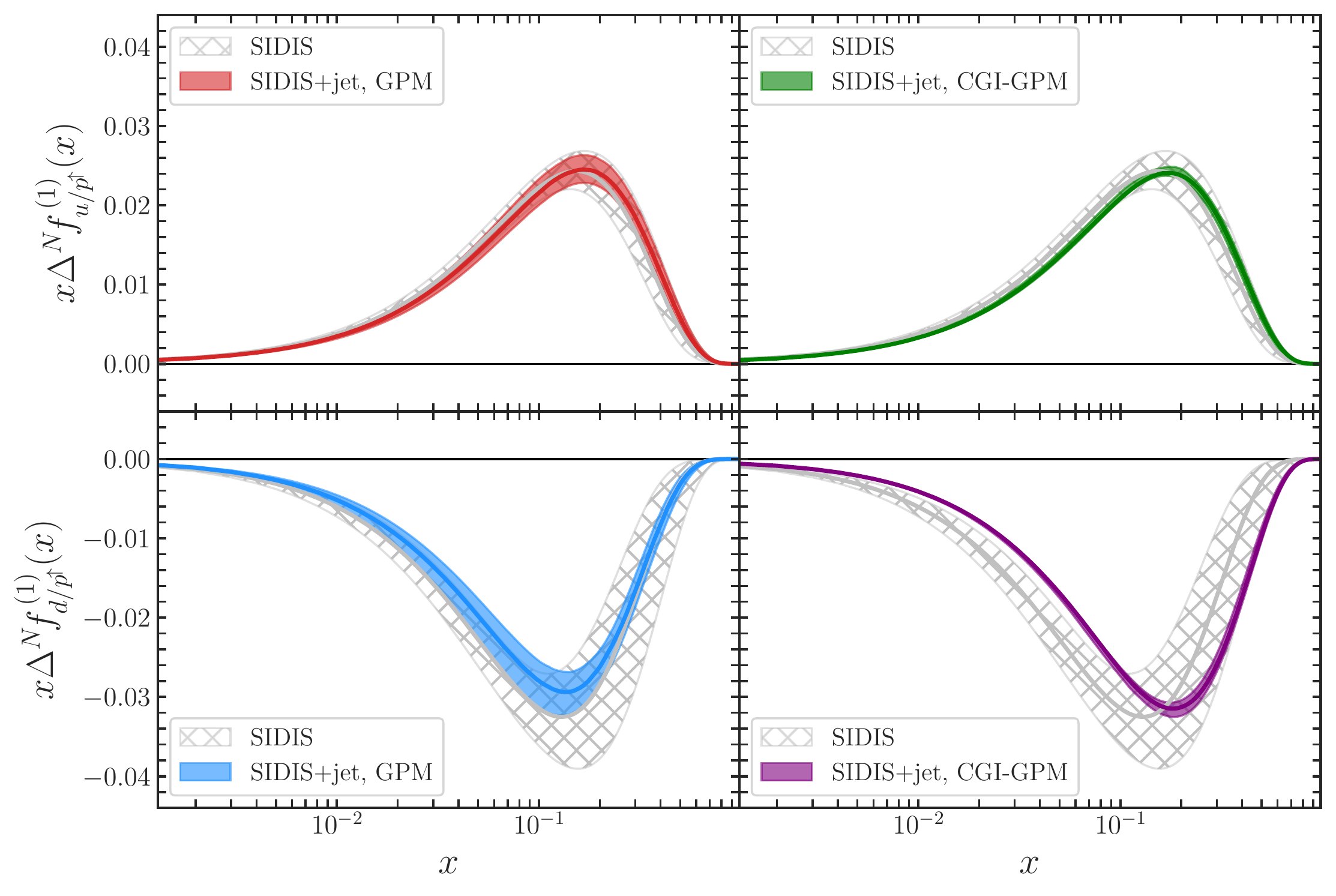} 
\caption{
Left: Transverse single-spin asymmetry as a function of $x_{F}$ for electromagnetic jets in transversely polarized proton-proton collisions at $\sqrt s=$ 200 and 500~GeV~\cite{STAR:2020nnl}.
Right: Comparison between the Sivers first $k_\perp$-moments from SIDIS data and their reweighted SIDIS+jet (data from STAR) in two frameworks: GPM and CGI-GPM \cite{Boglione:2021aha}. 
}
    \label{fig:EMjet_AN}
\end{figure*}

Firstly, the topological dependence of the $\pi^0$ $A_N$ was studied, and the $A_N$ of the isolated $\pi^0$'s (meaning no other particles around) are significantly larger than the non-isolated ones, as shown in the middle panel of Fig. \ref{fig:pi0_AN}. Consistent results were obtained in both $pp$ and $pA$ collisions with very weak $A$ dependence in $pA$~\cite{STAR:2020nnl,STAR:2020grs}.
This triggered discussions on the possible contribution from the diffractive process, which motivated a measurement of $A_N$ for singly and double diffractive events, utilizing the STAR Roman Pot detectors to tag diffractive processes with  scattered protons close to the beamline. Figure~\ref{fig:diffractive EM-jet AN} shows the preliminary results for forward diffractive EM-jet $A_{N}$ as a function of $x_{F}$ at $\sqrt{s}=$ 200 GeV~\cite{STAR:DiffractiveDIS2022}. The results favor a non-zero negative $A_{N}$ with 3.3$\sigma$ significance, so these diffractive processes are most probably not the source of the large positive $A_N$ of $\pi^0$. The negative contribution from diffractive jets is not currently described by theory.
%In addition, in the $pA$ collisions, no significant nuclear $A$ dependence was observed for $\pi^0$ $A_N$.

In studying the contribution from the final-state effect, STAR also measured the Collins asymmetry of $\pi^0$ in an electromagnetic jet, which is shown in the right panel of Fig.~\ref{fig:pi0_AN}.  The measured Collins asymmetry was consistent with zero, in agreement with a theoretical prediction based on collinear twist-3 factorization, resulting from significant cancellation between Collins effects of different quark flavors~\cite{Kang:2017btw}.

In a closely related study, RHICf has measured $A_N$ for neutral pions in 510 GeV $pp$ collisions at very large pseudorapidity ($\eta > 6$), very large $x_F$ (up to 0.8), and $p_T < 1$ GeV/$c$ \cite{RHIC-f:2020koe}.  The asymmetries that they found are similar to those at comparable $x_F$ and much higher $p_T$, as shown in the left panel of Fig.~\ref{fig:pi0_AN}.  A very recent calculation \cite{Kim:2022zgz} based on diffractive triple Regge exchange provides a very good description of the RHICf $A_N$ results.

Another study is the measurement of the $A_N$ for inclusive electromagnetic jets, which is considered only related to the initial-state effect.  
The results of electromagnetic jet $A_N$ in both 200 and 500 GeV $pp$ collisions are shown in the left panel of Fig.~\ref{fig:EMjet_AN}.
The electromagnetic jet $A_N$ was found to increase with $x_F$, but the magnitude is much smaller than the $\pi^0$ $A_N$.  These data have been included in the recent global fit of the Sivers function~\cite{Boglione:2021aha}, and showed a significant impact in constraining the Sivers function, as shown in the right panel of Fig.~\ref{fig:EMjet_AN}.

With the implementation of the Forward Upgrade at STAR in the $pp$ and $pA$ running in 2022 and 2024, we will be able to perform measurements with full jets in the forward rapidity region and also the Collins asymmetry with charge separated hadrons, which will provide a much deeper understanding of the underlying QCD mechanism. 

\subsection{Transverse single-spin asymmetry of weak bosons}

Proton-proton collisions at $\sqrt{s}$ = 510~GeV allow STAR to study the evolution and sign change of the Sivers function with weak bosons at mid-rapidity ($-1<y^{W^{\pm}/Z^{0}}<1$). By focusing on interactions in which the final state involves only leptons, and hence the transverse partonic motion must be in the initial state, one can test the predicted sign change in $A_N$ relative to interactions in which these terms must appear in the final state, such as SIDIS measurements. Following the low statistics proof-of-principle measurement using the 2011 data, STAR measured the transverse single-spin asymmetry $A_{N}$ for $W$ and $Z$ with 2017 data, which had about 14 times more integrated luminosity.

\begin{figure*}
    \centering
\includegraphics[width=4.8cm,clip]{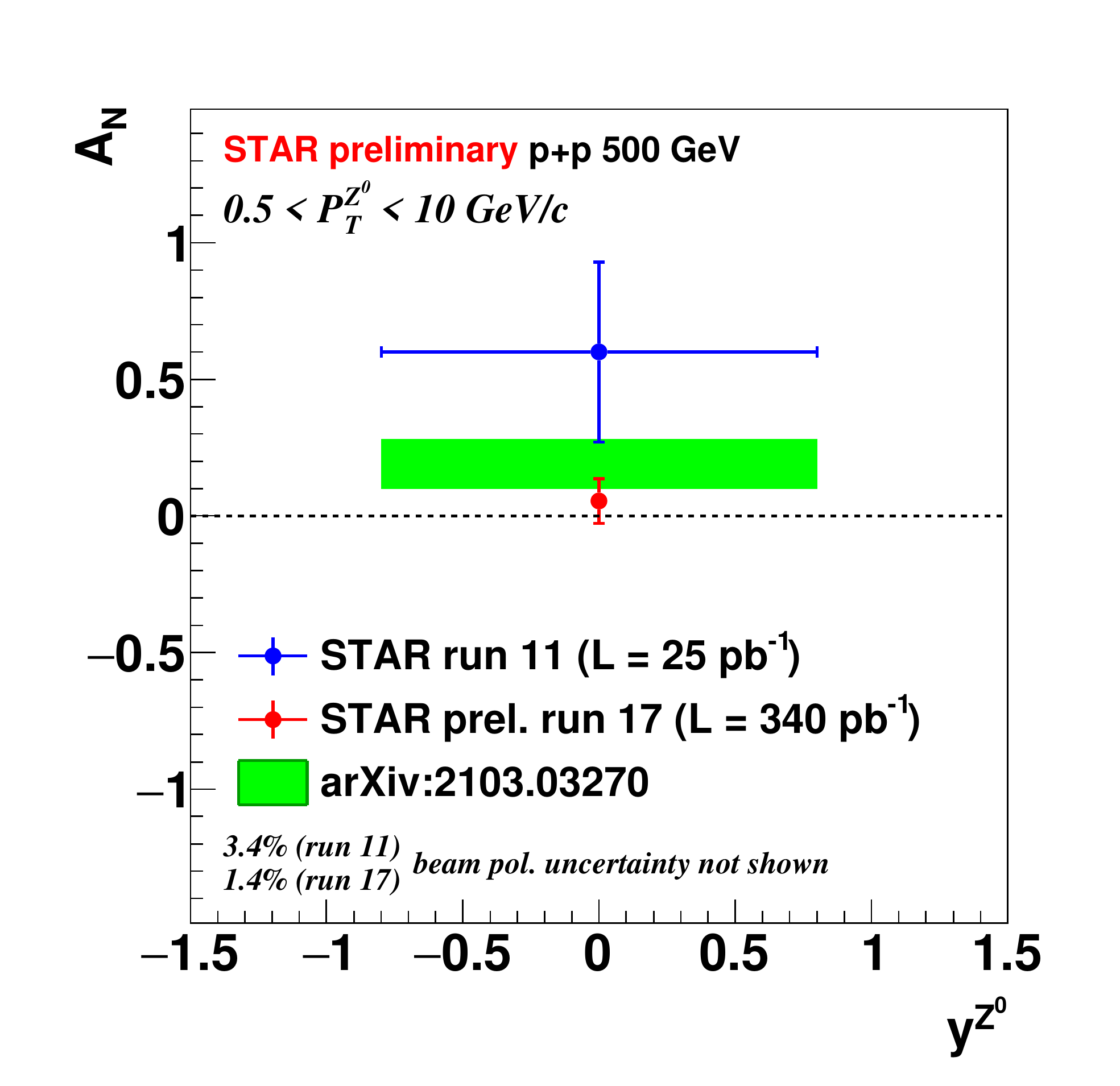}
\includegraphics[width=11.5cm,clip]{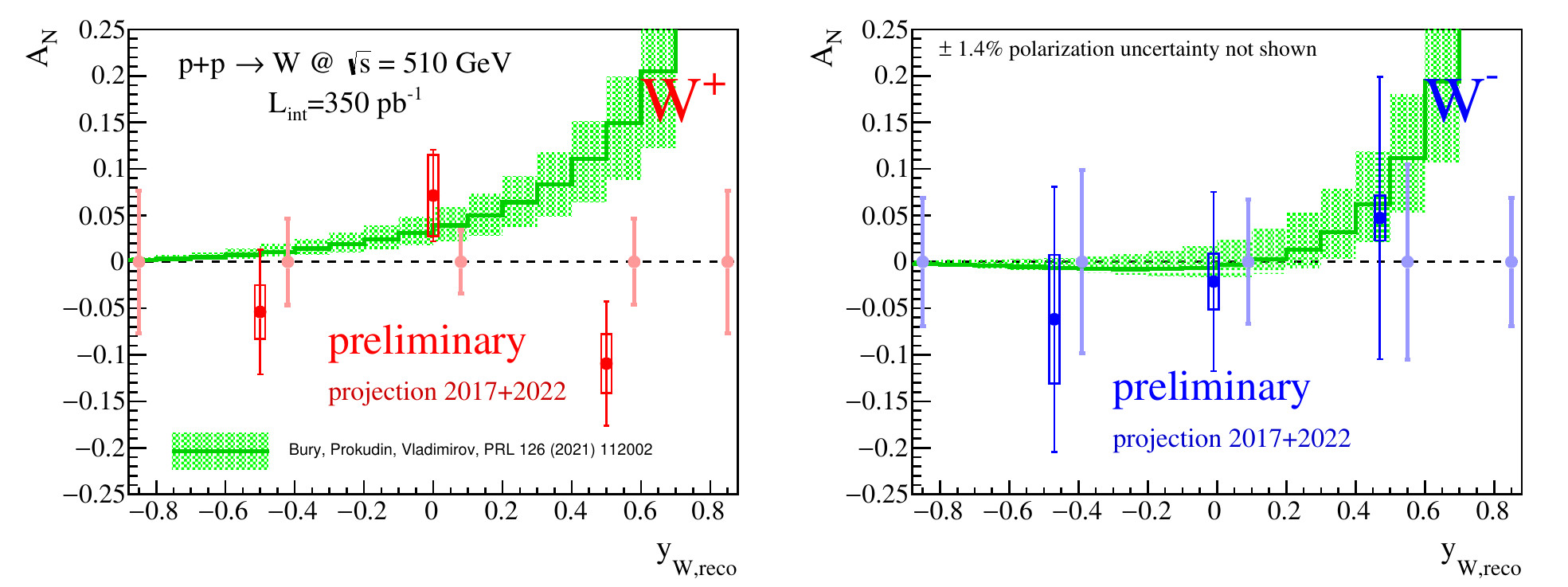} 
\caption{
Left: Transverse single-spin asymmetry of $Z^{0}$ from STAR 2011 and 2017 data. The results are compared with the calculation from \cite{Bury:2021sue}.
Middle and Right: Transverse single-spin asymmetry of $W^{\pm}$ from STAR 2017, and the projected statistical uncertainties from 2017 and 2022 data. The results are compared with calculation from \cite{Bury:2021sue} based on the next-to-next-to-next-to leading log ($\mathrm{N}^{3}$LL) accuracy TMD evolution from \cite{Bury:2020vhj}.
}
    \label{fig:plot_WZ_data}
\end{figure*}

\begin{figure*}
    \centering
\includegraphics[width=12cm,clip]{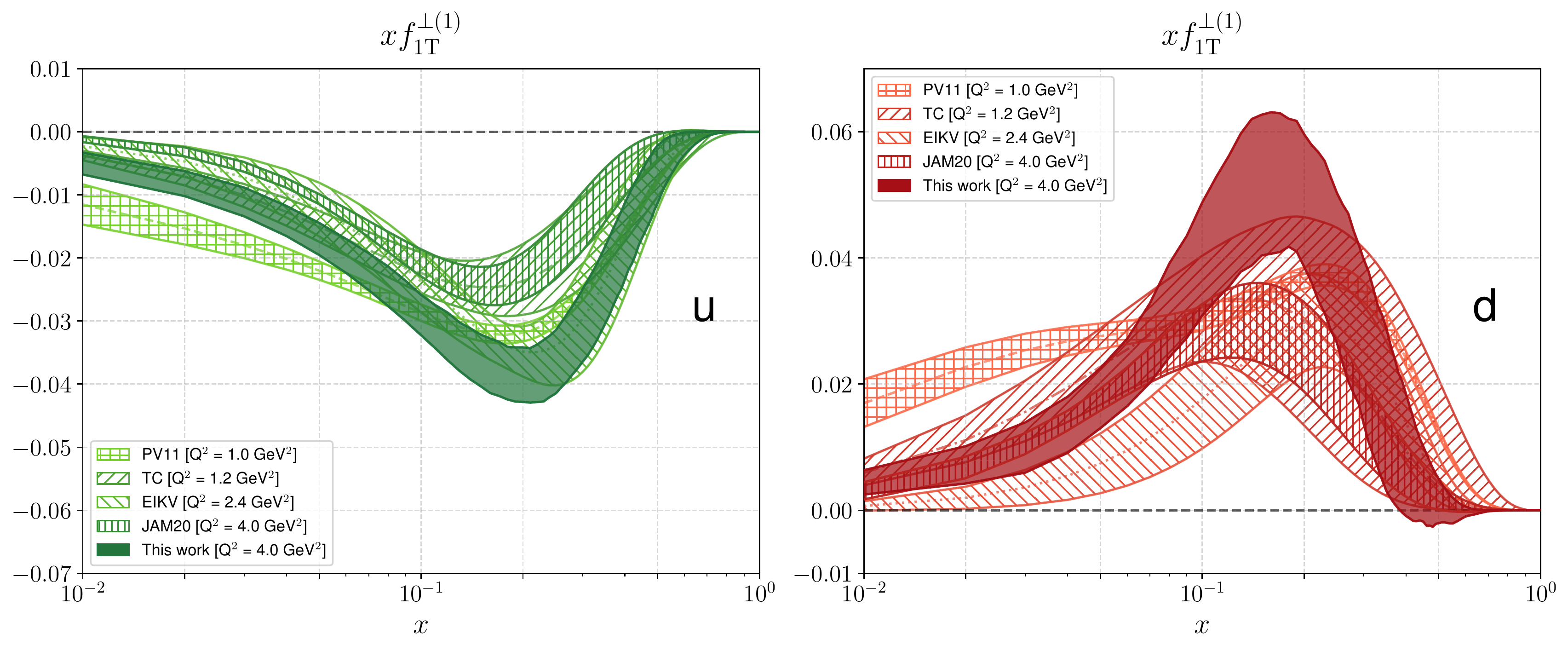} 
\caption{
The first transverse moment $x f_{1T}^{\perp (1)}$ of the Sivers
  TMD as a function of $x$ for the up (left panel) and down quark (right
  panel) extracted from world data including STAR 2011 $W/Z$ data. Solid band: the 68\% confidence interval obtained in
  this work at $Q^2 = 4$ GeV$^2$. The plot is from \cite{Bacchetta:2020gko}.}
    \label{fig:WZ_AN_impact_PV}
\end{figure*}

In Fig.~\ref{fig:plot_WZ_data}, the recent preliminary results on $A_{N}$ of $W/Z$ are compared with predictions from \cite{Bury:2021sue,Bury:2020vhj} that include STAR 2011 data. The recent global QCD extraction of the Sivers function including STAR 2011 $W$ and $Z$ $A_{N}$ data from \cite{Bacchetta:2020gko} can be found in Fig. \ref{fig:WZ_AN_impact_PV}. With the increased precision provided by Run-17, we find smaller asymmetries than were suggested by Run-11.
As a result, the increased statistics of the 2022 dataset are critical to improve the precision of our asymmetry measurements in order to provide a conclusive test of the Sivers' function sign change. Projected statistical uncertainties of $W$ $A_{N}$ from combined 2017 and 2022 data can be found in Fig. \ref{fig:plot_WZ_data}. The figure also illustrates that the improved tracking capabilities provided by the STAR iTPC upgrade will allow us to push our mid-rapidity $W^{\pm}$ and $Z$ measurements to larger rapidity $y^{W/Z}$, a regime where the asymmetries are expected to increase in magnitude and the anti-quark Sivers’ functions remain largely unconstrained.

\subsection{Transverse single-spin asymmetries of direct photons and heavy flavor decay leptons}
\begin{SCfigure}[][h]
    \centering
    \hspace{-0.8cm}
    \includegraphics[width=0.55\textwidth]{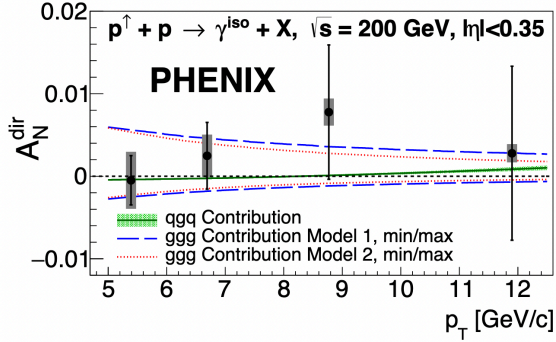}
    \caption{Transverse single-spin asymmetry of isolated direct photons at $\sqrt{s}=$ 200 GeV compared with calculations from $qgq$ and $ggg$ correlation functions. Source:~\cite{PHENIX:2021ph}.}
    \label{fig:AN_photon}
\end{SCfigure}

PHENIX has reported the first direct photon transverse single-spin asymmetry result at RHIC \cite{PHENIX:2021ph}. The asymmetry was measured at midrapidity $|\eta|<0.35$ in $pp$ collisions at $\sqrt{s}=$ 200 GeV. Photons do not interact via the strong force, and at this kinematics they are produced dominantly by the quark-gluon Compton process. Therefore, the measurement offers a clean probe of gluon dynamics that is only sensitive to initial-state effects.
The asymmetry is shown in Fig.~\ref{fig:AN_photon} and is consistent with zero to within 1\% across the measured $p_{_{T}}$ range. The result is also compared with predictions from collinear twist-3 correlation functions. 
The solid green curve shows the contribution from $qgq$ correlation function \cite{Koike:2015qgq} while the dashed (blue) and dotted (red) curves are from $ggg$ correlation functions \cite{Koike:2012ggg}. Given the small predicted contributions from $qgq$ correlation functions to the asymmetry, the result can provide a constraint on the $ggg$ correlation function. sPHENIX is expected to significantly improve direct photon $A_N$ measurements shrinking the uncertainties by more than a factor of two. 

\begin{SCfigure}[][h]
    \centering
    \includegraphics[width=0.5\textwidth]{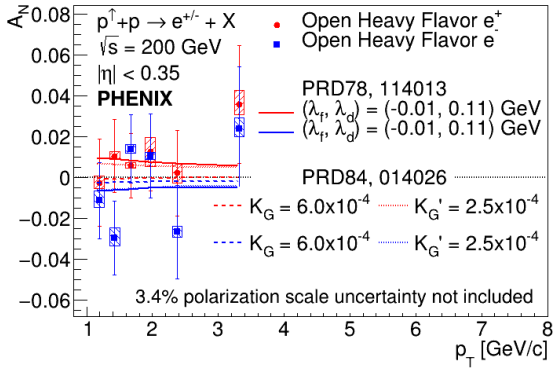}
    \caption{Transverse single-spin asymmetries of heavy flavor decay electrons at $\sqrt{s}=$ 200 GeV \cite{PHENIX:2022znm} including parameterizations of the tri-gluon correlator in two theoretical models and the best values fitting the data \cite{Kang:2008ih,Koike:2011mb}.}
    \label{fig:AN_HFe}
\end{SCfigure}

%RCS
Similarly, the production of open heavy flavor at RHIC energies is dominated by gluon-gluon hard interactions. As such, also in single-spin asymmetries of heavy flavor decay leptons no final-state effect contributions are expected, and one is almost entirely sensitive to the initial state effects of the gluon correlators. 
The recent heavy flavor decay electron single-spin asymmetries at central rapidities obtained at PHENIX \cite{PHENIX:2022znm} are the first that quantify the gluon correlator contributions in two theoretical models \cite{Kang:2008ih,Koike:2011mb}, as can be seen in Fig.~\ref{fig:AN_HFe}. While each decay lepton asymmetry is only sensitive to a linear combination of the two model parameters, the combination of both charges enables the determination of both. In the 2024 data taking period, these measurements can be augmented by sPHENIX measurements that reconstruct D mesons directly and are expected to provide even higher precision to the tri-gluon correlator. 

\subsection{Nuclear dependence of single spin asymmetries}
\begin{SCfigure}[][h]
    \centering
    \includegraphics[width=0.5\textwidth]{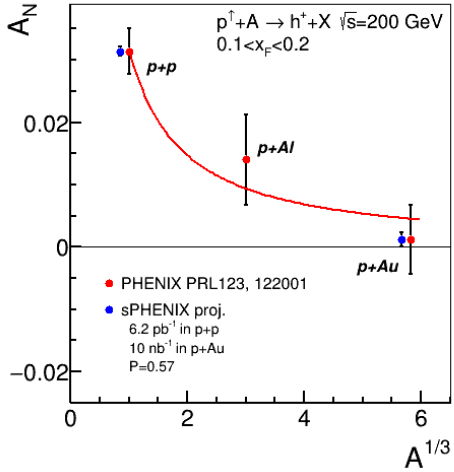}
    \caption{$A$ dependence of transverse single-spin asymmetries of positively charged hadrons at $\sqrt{s}=$ 200 GeV at pseudo-rapidities of 1.2 to 2.4 measured at PHENIX \cite{PHENIX:2019ouo}, and sPHENIX projected uncertainties for data to be collected with streaming readout.}
    \label{fig:Adep}
\end{SCfigure}

In 2015, RHIC also investigated polarized proton-nucleus collisions with either Al or Au beams. These have been utilized to study the $A$ dependence of the nonzero single-spin asymmetries that were observed for hadrons in the forward region. In PHENIX the asymmetries for charged hadrons at rapidities of 1.2 to 2.4 were studied. A strong nuclear dependence was observed that was consistent with an $A^{-1/3}$ suppression for positive hadrons \cite{PHENIX:2019ouo}, as shown in Fig.~\ref{fig:Adep}. A similar suppression is also seen as a function of the centrality of the collisions.  

\begin{SCfigure}[][h]
    \centering
    \includegraphics[width=0.5\textwidth]{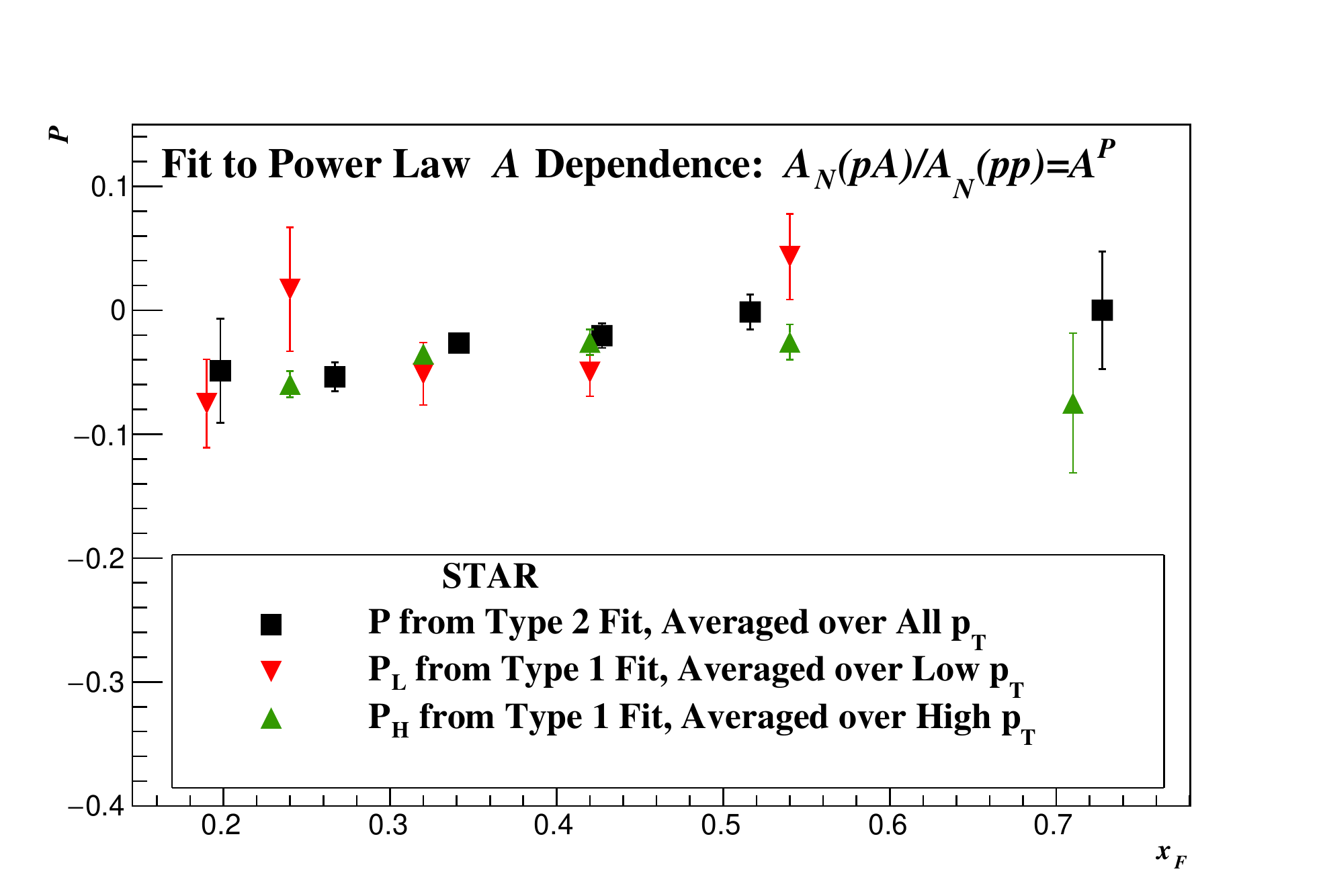}
    \caption{The exponent, $P$, for nuclear $A$ dependence of the $\pi^0$ transverse single-spin asymmetry ratio of $pA$ to $pp$ as a function of $x_F$ at $\sqrt{s}=$ 200 GeV at $2.7<\eta<3.8$ at STAR \cite{STAR:2020grs}. The main difference of two types of fits is with and without correlated uncertainties.}
    \label{fig:Adep-pi0}
\end{SCfigure}

STAR has also published the $A$ dependence for neutral pions at forward rapidities of 2.7 to 3.8 and higher $x_F$ that also show a suppression of the asymmetries \cite{STAR:2020grs}. However, in that rapidity region the suppression appears much smaller than seen by PHENIX, as seen in Fig.~\ref{fig:Adep-pi0}. 
The initial motivation for studying the nuclear dependence of the single-spin asymmetries originates from possible saturation effects on these asymmetries, but it has since been realized that the presented measurements neither reach $x$ nor scales that are low enough for such effects to be relevant \cite{Benic:2018amn}. As such, there is at present no clear understanding of the mechanism that produces the suppression of these asymmetries. 

The data to be collected by sPHENIX in $pp$ and $pAu$ collisions would not only considerably improve the precision of PHENIX measurements for charged hadron TSSA (Fig.~\ref{fig:Adep}), but also allow for fine binning of TSSA in $p_T$ and $x_F$ in extended ranges. That will provide invaluable information for studying rich phenomena behind TSSA in hadronic collisions, utilizing RHIC's unique capabilities to collide high energy polarized protons and heavy
nuclei.

In the far forward region also the nuclear dependence of neutron asymmetries was extracted as a function of transverse momentum and the longitudinal momentum fraction \cite{PHENIX:2017oku,PHENIX:2021yxs}. Neutron asymmetries in proton-proton collisions can be described by the interference of pion and other meson interactions between the two colliding nucleons \cite{Kopeliovich:2011bx} and are found to be negative. In contrast, the $pAl$ asymmetries are on average close to zero, while the $pAu$ asymmetries change sign and have a significantly larger magnitude. It was found that the origin of this nuclear dependence originates from the additional contribution of ultra-peripheral collisions that increase quadratically with the charge of the nucleus \cite{Mitsuka:2017czj}. When correlating the asymmetries with event activity related to hadronic activity, one indeed sees that the asymmetries remain negative while the events more likely to originate from ultra-peripheral collisions show even larger, positive asymmetries already for $pAl$ collisions.  

% Bill and Kong
\subsection{Ultra-peripheral collisions}

%The formalism of generalized parton distributions (GPDs) provides a theoreticalframework which addresses some of the above questions~\cite{Mueller:1998fv,Ji:1996ek,Radyushkin:1996nd,Burkardt:2000za}.
Constraints on GPDs have mainly been provided by exclusive reactions in  deep inelastic scattering (DIS), e.g. deeply virtual Compton scattering. RHIC, with its unique capability to collide transversely polarized protons at high energies, has the opportunity to measure $A_N$ for exclusive $J/\psi$ production in
ultra-peripheral collisions (UPCs)~\cite{Klein:2003qc}. In such a UPC process, a photon emitted by the opposing beam particle ($p$ or $A$) collides with the polarized proton. The measurement is at a fixed $Q^2 \sim M_{J/\psi}^2 \approx 10$ GeV$^2$ and $10^{-4} < x < 10^{-1}$. A nonzero asymmetry would be the first signature of a nonzero GPD $E_g$ for gluons, which is sensitive to spin-orbit correlations and is intimately connected with the orbital angular momentum carried by partons in the nucleon and thus with the proton spin puzzle.

\begin{figure*}[tbh]
\centering
\includegraphics[width=0.46\textwidth]{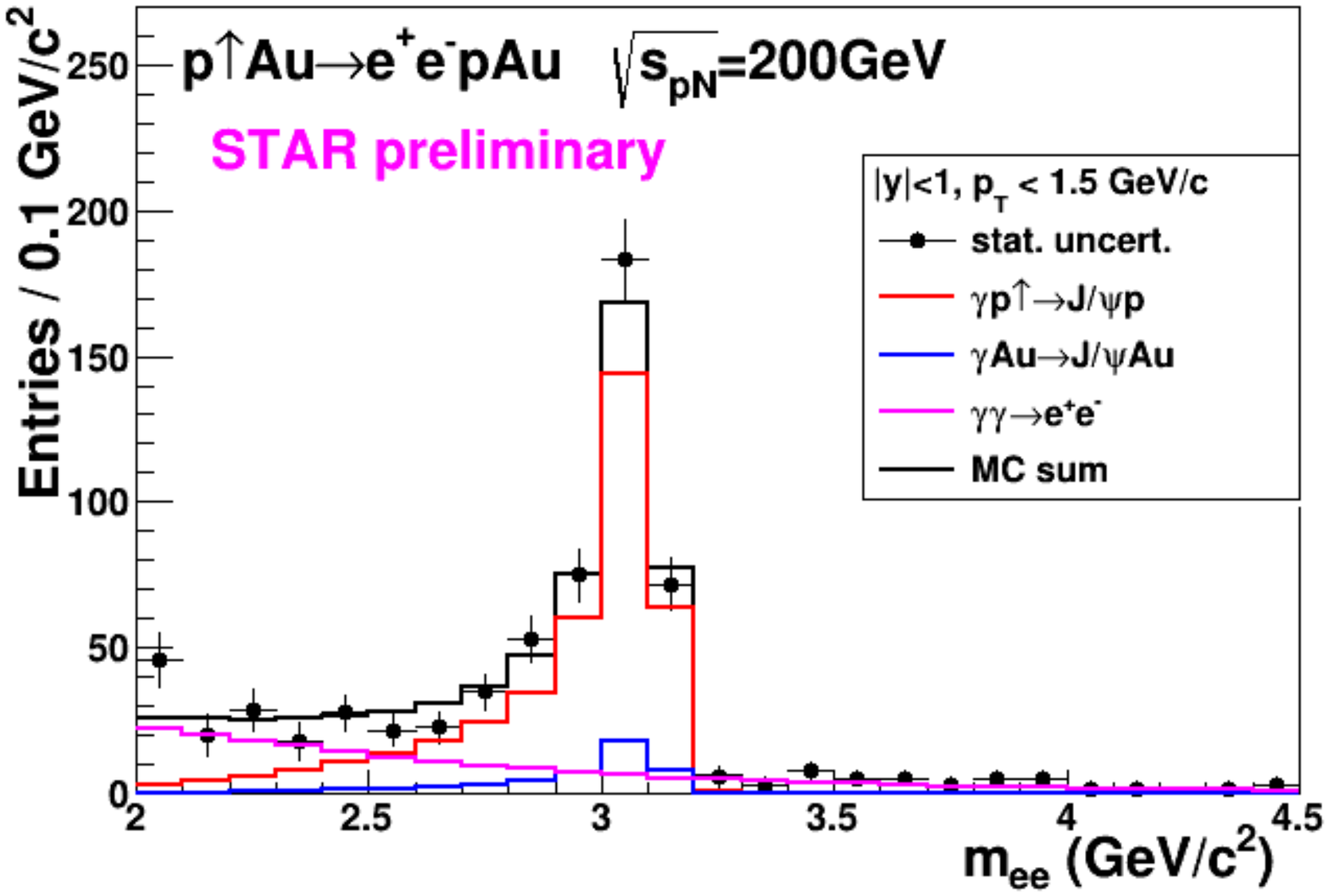}
\includegraphics[width=0.46\textwidth]{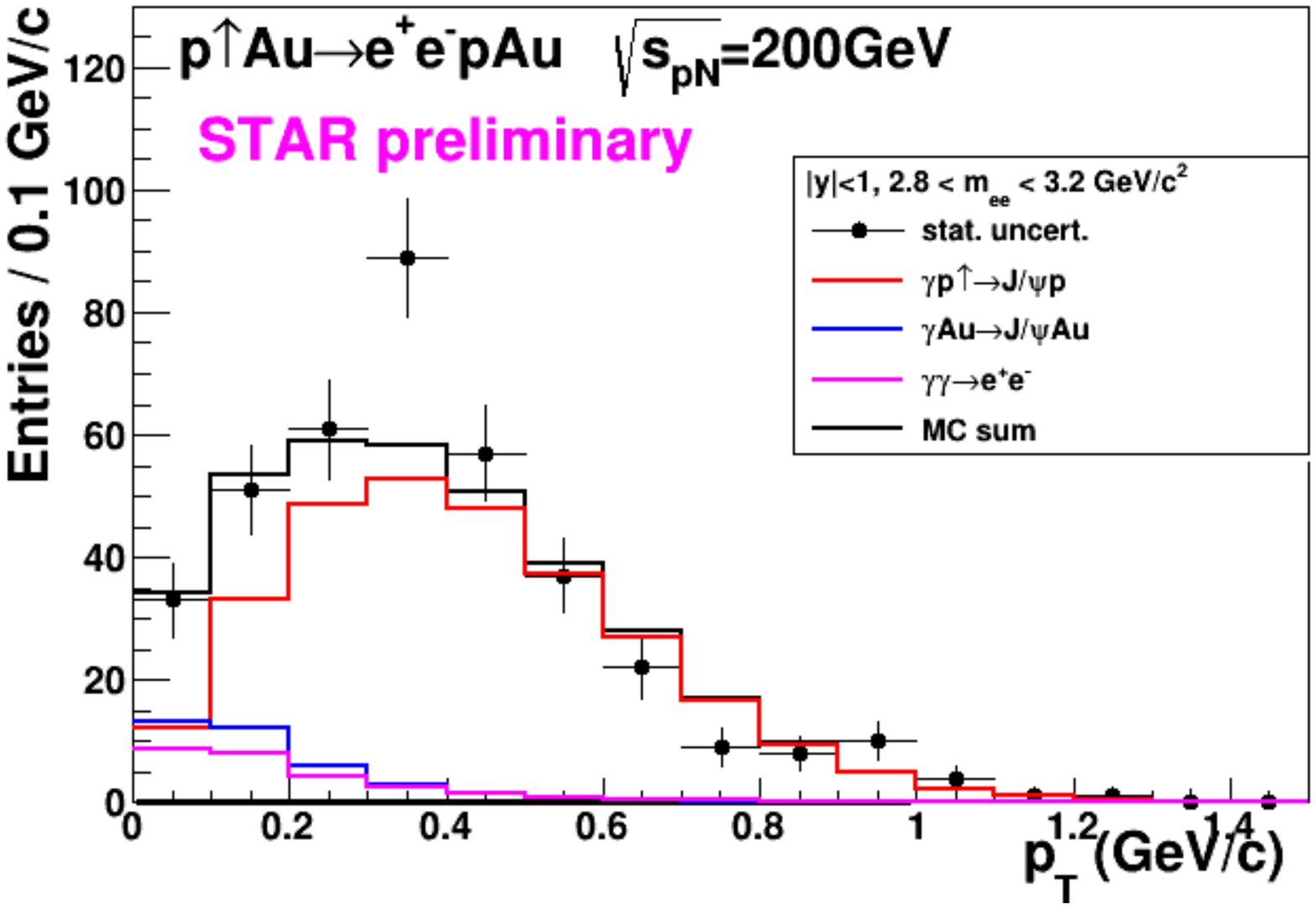}
\caption{Mass distribution of selected $e^+e^-$ pairs (left), and $p_T$ distribution of the $J/\psi$ mass peak (right). The colored histograms are the indicated processes modelled by STARlight and the sum fit to the data.}
\label{fig:JPsiMpT}
\end{figure*}

\begin{figure*}[tbh]
\centering
\includegraphics[width=0.46\textwidth]{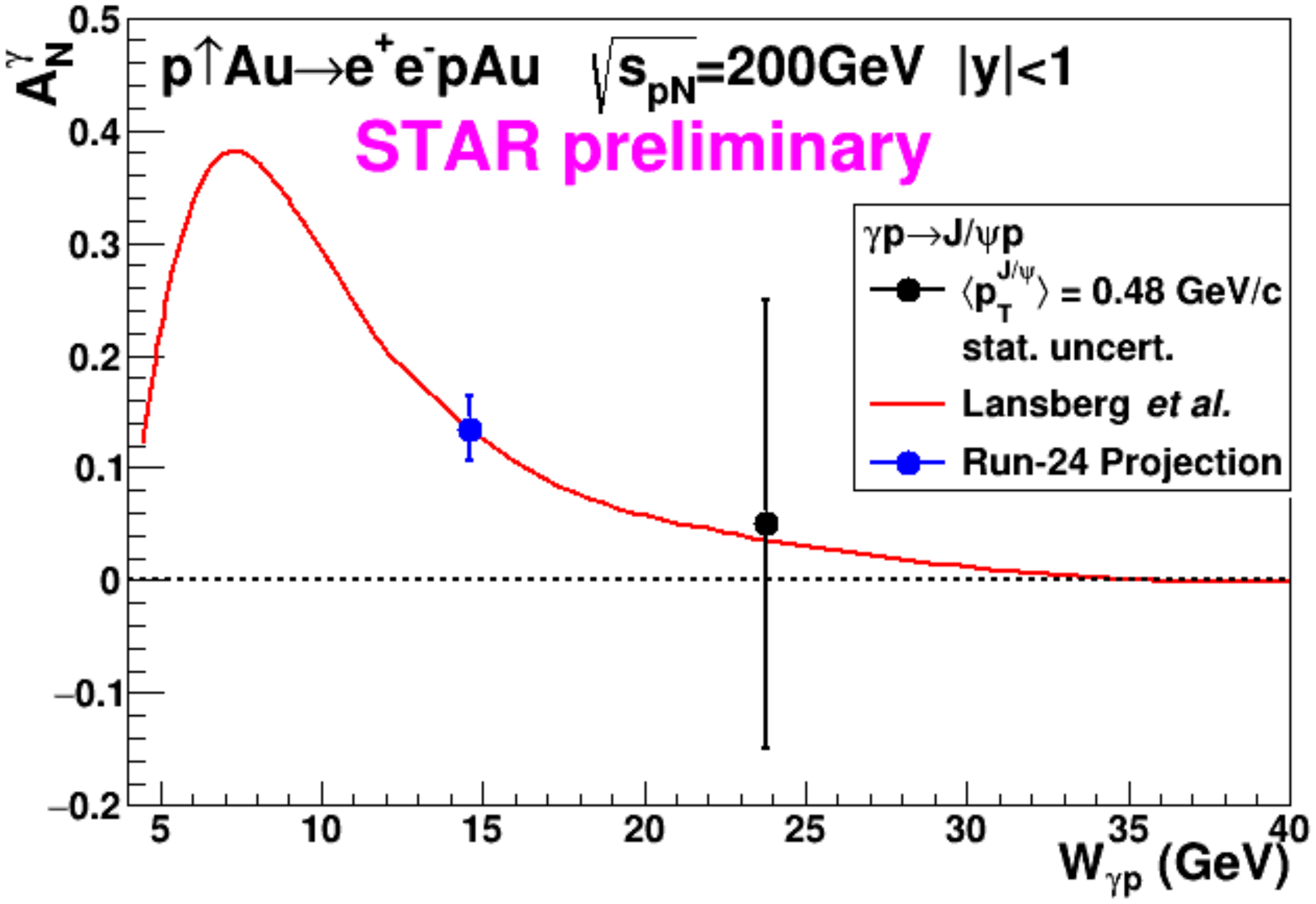}
\includegraphics[width=0.46\textwidth]{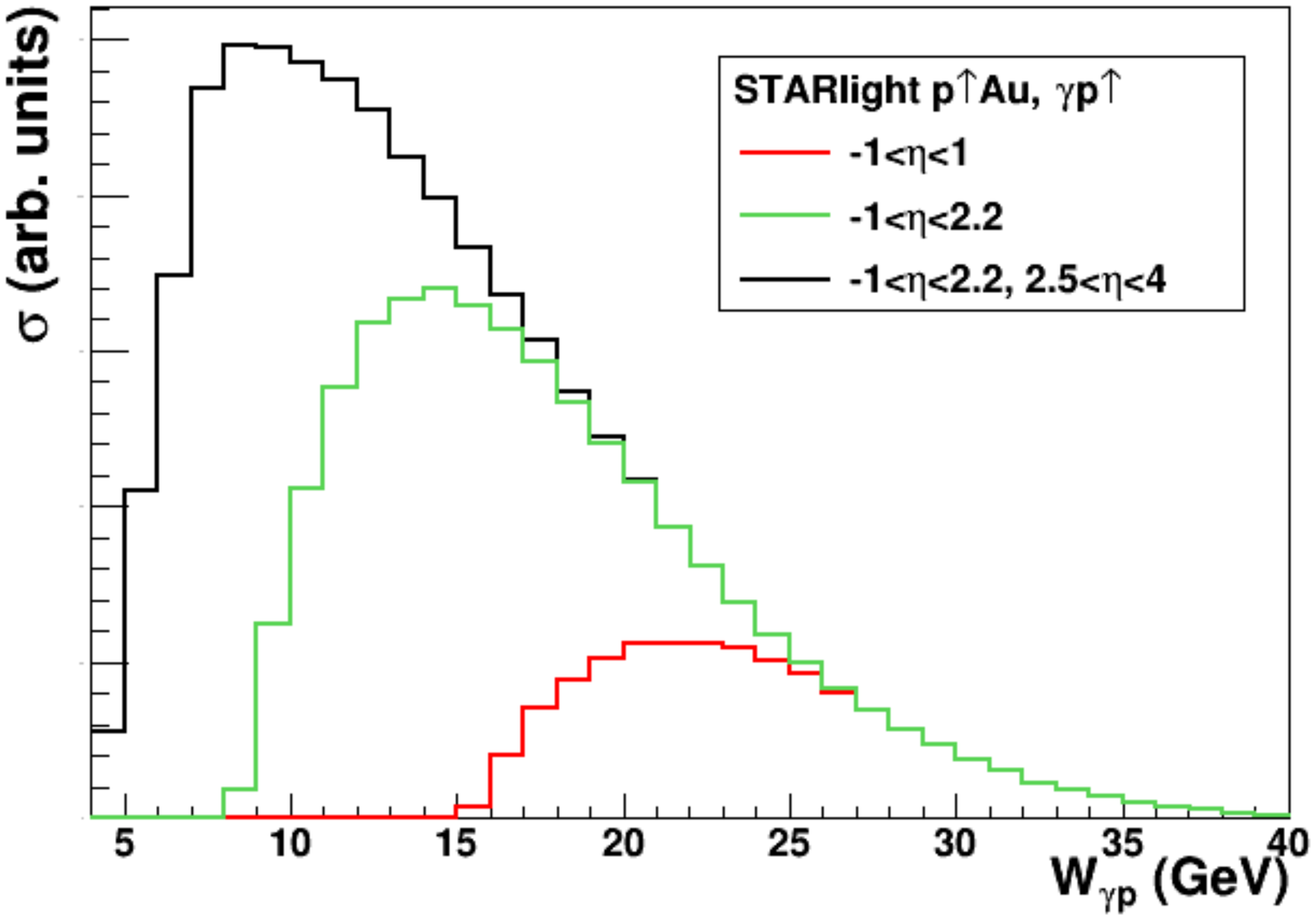}
\caption{ Left: The measured $J/\psi$ transverse asymmetry $A_N^{\gamma}$
and a prediction based on a parameterization of $E_g$.
Right: The accepted cross section for $\gamma+p^\uparrow \rightarrow J/\psi$
for various detector pseudorapidity $\eta$ ranges; the black curve shows the result for the full STAR detector with the Forward Upgrade and the iTPC.}
\label{fig:JPsiANsigma}
\end{figure*}

The Run-15 $p^{\uparrow}Au$ data allowed a proof-of-principle of such a measurement.
A trigger requiring back-to-back energy deposits in the Barrel Electromagnetic Calorimeter selected $J/\psi$ candidates. The $e^+e^-$ mass distribution after selection cuts is shown in the left of Fig.~\ref{fig:JPsiMpT}, and the pair $p_T$ distribution of the $J/\psi$
mass peak is shown on the right of that figure.
The data are well described by the STARlight model~\cite{Klein:2016yzr} (colored histograms in the figure), including the dominant $\gamma$+$p^\uparrow \rightarrow J/\psi$ signal process and the
$\gamma$+$Au\rightarrow$$J/\psi$ and $\gamma$+$\gamma$$\rightarrow$$e^+e^-$ background processes.
The left of Fig.~\ref{fig:JPsiANsigma} shows the STAR preliminary measurement (solid circle marker) of the transverse asymmetry $A_N^{\gamma}$ for the $J/\psi$ signal, which has a mean photon-proton
center-of-mass energy $W_{\gamma p} \approx 24$ GeV. The result is consistent with zero. Also shown is a prediction based on a parameterization of $E_g$~\cite{Lansberg:2018fsy}; the present data provide no discrimination of this prediction.

This measurement can be greatly improved with a high statistics transversely polarized $p^{\uparrow}Au$ Run-24. The integrated luminosity for the Run-15 measurement was 140 nb$^{-1}$; Run-24 will provide about 1.2 pb$^{-1}$, allowing a sizeable reduction of statistical uncertainty in the same $W_{\gamma p}$ range. In addition, the Forward Upgrade and iTPC will provide a significant extension of the $W_{\gamma p}$ range of the measurement. The right panel of Fig.~\ref{fig:JPsiANsigma} shows the accepted cross section for $\gamma$+$p^\uparrow \rightarrow J/\psi$ for various detector pseudorapidity ranges. With the full detector, the sensitive cross section is a factor of five times the central barrel alone.
Also, the accepted region has a lower mean $W_{\gamma p} \approx 14$ GeV. Predictions based on $E_g$ parameterizations such as shown in the figure have a larger asymmetry at lower $W_{\gamma p}$, with increased possibility of a nonzero result. 
%and the expected asymmetry is substantially larger. 
The projected statistical uncertainty on $A_N^{\gamma}$ is shown in the left of Fig.~\ref{fig:JPsiANsigma} (blue square marker), offering a powerful test of a non-vanishing $E_g$. Alternatively, the increased statistics will allow a measurement of $A_N^{\gamma}$ in bins of $W_{\gamma p}$.

The UPC cross section scales with $\sim Z^2$ of the the nucleus emitting the photon; for protons this is $1/79^2$ relative to Au nuclei, which makes analogous measurements in \pp\ collisions extremely luminosity-hungry. Therefore, the $pAu$ run is important for this measurement.

%\end{multicols}

%% file: Appendix.tex
\subsection{STAR Forward Upgrade}
\begin{figure}[h]
\centering
\includegraphics[width=0.9\textwidth]{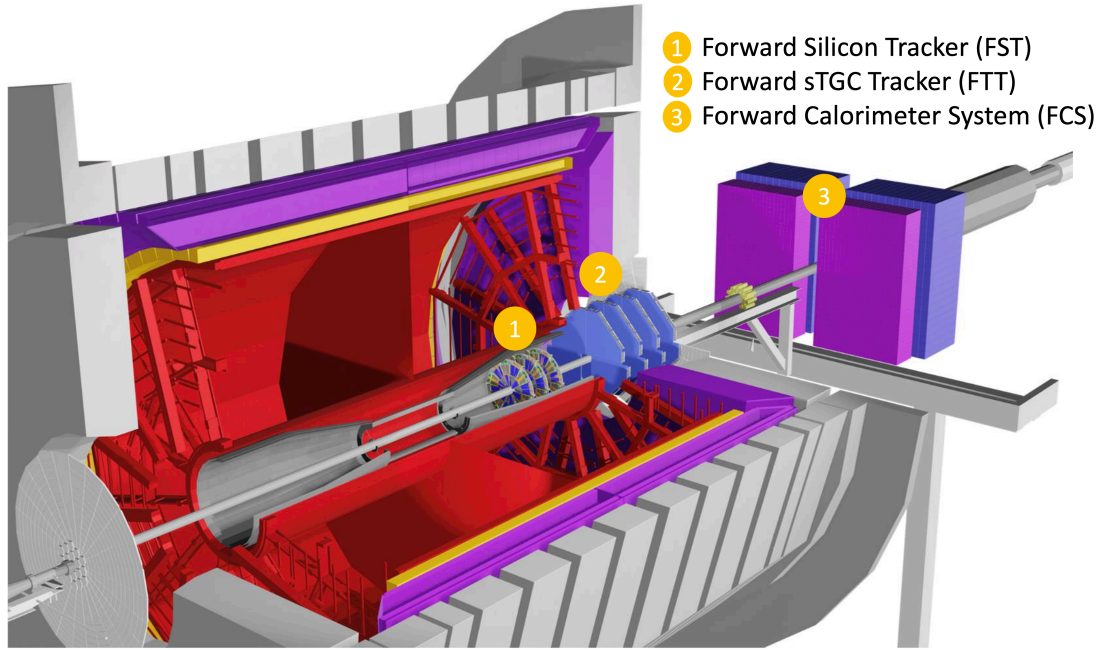}
\caption{STAR detector with Forward Upgrades}
\label{fig:forwardupgrade}
\end{figure}

The STAR forward upgrade consists of four major new subsystems, an electromagnetic calorimeter, a hadronic calorimeter and a tracking system, formed from a silicon detector and a small-strip Thin Gap Chambers tracking detector. It has superior detection capabilities for neutral pions, photons, electrons, jets, and leading hadrons within the pseudorapidity range 2.5 < $\eta$ < 4, see Fig. \ref{fig:forwardupgrade}. The construction of the electromagnetic and hadronic calorimeters was successfully completed by the end of 2020. They were fully installed, instrumented, and commissioned during the 2021 RHIC running period. The tracking detectors were installed in summer and fall 2021, on schedule and ready for the start of Run-22. Note that the entire construction, installation, and commissioning of the four systems were completed in the pandemic period. Enormous efforts were made to keep the forward upgrades on schedule. During Run-22, despite all the difficulties from the machine side, the forward upgrades performed exceptionally well and took data smoothly throughout the run. The forward upgrades will continue taking data in parallel with sPHENIX through Run-25.

\begin{itemize}
    \item Forward Calorimeter System:

The Forward Calorimeter System (FCS) consists of an Electro-Magnetic Calorimeter (Ecal) with 1486 towers, and a Hadronic Calorimeter (Hcal) with 520 towers. All SiPM sensors, front-end electronics boards and readout \& triggering boards called DEP were installed, commissioned and calibrated during Run-21. Signal splitter boards for the west EPD detector were installed before Run-22, and the west EPD was used as pre-shower detector in the electron triggers. FPGA code for FCS triggers was developed in fall 2021, and total of 29 triggers, including triggers for electrons, di-electrons, jets, di-jets, hadrons, and photons were commissioned and verified within a few days of RHIC starting to deliver stable \pp\ collisions, and then used for data taking throughout Run-22 successfully. 
FCS operations during Run-22 were successful and smooth. The only minor exceptions were 3 low-voltage power supply modules needing to be replaced, and occasional power cycling of electronics being needed due to beam related radiation upsets in the electronics. All 1486 channels of Ecal  worked with no bad channels, and the Hcal had only a couple of dead channels. Radiation damage to the SiPM sensors due to beam was within expectations. There was an unexpected loss of signal amplitudes of $\sim$20\% per week in the Ecal near the beam, which turned out to be radiation damage in the front-end electronics boards. The loss of signal was compensated during Run-22 by changing the gain factors on the DEP boards, attenuator settings in the front-end electronics, and raising  the voltage settings tower by tower based on LED signals. Details of the radiation damage on the front-end electronics are currently under investigation.

 \item Small-strip Thin Gap Chambers:
 
The sTGC has four identical planes, each plane has four identical pentagonal shaped gas chambers. These gas chambers are made of double-sided and diagonal strips that give $x$,$y$,$u$ in each plane.  Sixteen chambers and about 5 spare chambers were built at Shandong University in China. A custom designed and fabricated aluminum frame allowed to fit the detector inside the pole-tip of the STAR magnet and around the beam-pipe on the west side of STAR. The sTGC chambers are operated with a quenching gas mixture of $n$-Pentane and CO$_2$ at a ratio of 45\%:55\% by volume at a typical high voltage of 2900 V. This gas mixture allowed the chambers to operate in a high amplification mode. 
The sTGC was fully installed prior to the start of Run-22, and the detector was fully commissioned during the first few weeks of the run. The operating point of the high voltage was scanned for optimum efficiency. The gas chambers were stable at the desired operational high voltage and at the high luminosity, also the leakage current was well within the operational limits. In-house, a newly designed and built gas system for mixing, and supplying the gas along a long-heated path to deliver to the chambers, met the above requirements, and performed exceptionally well during Run-22.

\item Forward Silicon Tracker:

The Forward Silicon Tracker (FST) consists of three identical disks, and each disk contains 12 modules. Each module has 3 single-sided double-metal Silicon mini-strips sensors which are read out by 8 APV chips. The module production was done by NCKU, UIC, and SDU. The readout was done by BNL and IU. The cooling was provided by NCKU and BNL. The installation of the FST was completed on August 13th, 2021, and the first \pp\ 510 GeV collision data were recorded on December 15, 2021. The FST ran smoothly through the whole Run-22, and the detector operation via slow control software was minimal to the shift crew.
\end{itemize}

\begin{figure}
    \centering
    \includegraphics[width=0.9\textwidth]{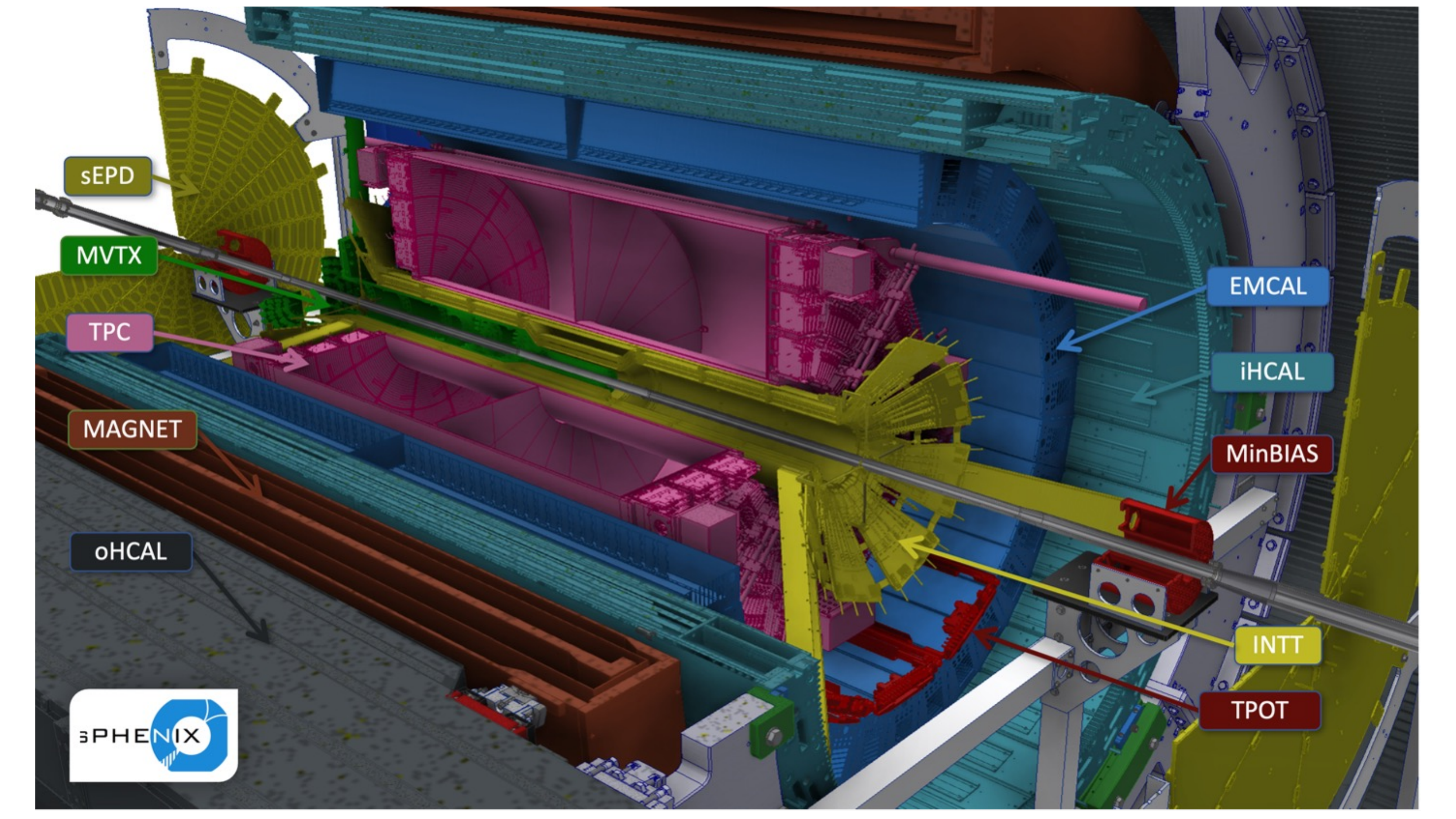}
    \caption{sPHENIX detector layout.}
    \label{fig:sPHENIX}
\end{figure}

\subsection{sPHENIX Detector}

sPHENIX is a major upgrade to the PHENIX experiment at RHIC capable of measuring jets, photons, charged hadrons, and heavy flavor probes. sPHENIX will play a critical role in the completion of the RHIC science mission, focused on the studies of the microscopic nature of Quark-Gluon Plasma. Polarized proton collisions as well as proton-nucleus collisions will also provide key opportunities for cold QCD measurements. 

sPHENIX is a central rapidity detector ($|\eta|<1.1$) built around the Babar solenoid with magnetic field up to 1.5T. The major systems are a high precision tracking system, and electromagnetic and hadronic calorimeters, see Fig.~\ref{fig:sPHENIX}.

The electromagnetic calorimeter is a compact tungsten-scintillating fiber design located inside the solenoid. The outer hadronic calorimeter consists of steel in which scintillator tiles with light collected by wavelength shifting fibers are sandwiched between tapered absorber plates that project nearly radially from the interaction point. It also serves as a flux return of the 1.5 T superconducting solenoid. The inner HCal is instrumented with scintillating tiles similar to the tiles used in the Outer HCal, and serves as a support structure of the electromagnetic calorimeter. The calorimeters use a common set of silicon photomultiplier photodetectors and amplifier and digitizer electronics. Based on test beam data, such a calorimeter system is expected to provide the energy resolution of  $\sigma_E/E=13\%/\sqrt{E[GeV]} \oplus 3\%$ for electromagnetic showers, and $\sigma_E/E=65\%/\sqrt{E[GeV]} \oplus 14\%$ for hadrons.

The central tracking system consists of a small Time Projection Chamber (TPC), micro vertex detector (MVTX) with three layers of Monolithic Active Pixel Sensors (MAPS), and two layers of the intermediate silicon strip tracker within the inner radius (INTT). Such a system provides momentum resolution $\sigma_{pT}/p_T<0.2\%\cdot p_T \oplus 1\%$ for $p_T=$0.2--40 GeV/c, and Distance of Closest Approach (DCA) resolved at 10 $\mu$m for $p_T>2$ GeV/c. The INTT with its fast integration time resolves beam crossings and provides pileup suppression. 

The other sPHENIX subsystems are the Minimum Bias Detector (MBD) consisting of the refurbished PHENIX Beam-Beam Counter, Event Plane Detector (sEPD) consisting of two wheels of scintillator tiles positioned at $2 < |\eta| < 4.9$ and serving for event plane measurements, and Micromegas-based TPC Outer Tracker (TPOT), offering calibration of beam-induced space charge distortions in TPC. 

High speed data acquisition system is designed to be capable of taking minimum bias $AuAu$ collisions at 15 kHz with greater than 90\% live time, and jet and photon triggers for $pp$ and $pA$ operation. The DAQ system is design to be capable to work in hybrid mode: along with triggered data it will collect a significant fraction ($\sim10\%$) of all collision data from tracking detectors in streaming readout regime, which will greatly extend physics program in $pp$
and $pAu$ running.